\documentclass[structabstract]{aa}  
 \usepackage{natbib}                                   
 \usepackage{upgreek}
\usepackage{graphicx}
\usepackage{amssymb}
\usepackage{txfonts}
\usepackage{color}
\begin{document}
   \title{A {\it Herschel} [C\,{\sc ii}] Galactic plane survey I: the global
   distribution of ISM gas components\thanks{{\it Herschel} is an ESA space observatory 
with science instruments provided by European-led Principal Investigator 
consortia and with important participation from NASA.}}

   \author{J. L. Pineda,  W. D. Langer, T. Velusamy, and P. F. Goldsmith}

\authorrunning{Pineda, Langer, Velusamy, and Goldsmith}
\titlerunning{A {\it Herschel} [C\,{\sc ii}]  Galactic Plane Survey}

   \institute{Jet Propulsion Laboratory, California Institute of Technology, 4800 Oak Grove Drive, Pasadena, CA 91109-8099, USA\\
              \email{Jorge.Pineda@jpl.nasa.gov}
             }
   \date{Received January 30, 2013; accepted April 11, 2013}

 
  \abstract{ The [C\,{\sc ii}] 158$\mu$m line is an important tool for
understanding the life cycle of interstellar matter. Ionized carbon is
present in a variety of phases of the interstellar medium (ISM),
including the diffuse ionized medium, warm and cold atomic clouds,
clouds in transition from atomic to molecular, and dense and warm
photon dominated regions (PDRs).  }
{ 
Velocity--resolved observations of [C\,{\sc ii}] are the most powerful
technique available to disentangle the emission produced by these
components.  These observations can also be used 
to trace CO--dark H$_2$ gas and  determine  the total mass of the ISM.
}
{ The Galactic Observations of Terahertz C+ (GOT\,C+) project surveys
 the [C\,{\sc ii}] 158$\mu$m line over the entire Galactic disk with
 velocity--resolved observations using the \it Herschel\rm/HIFI
 instrument.  We present the first longitude--velocity maps of the
 [C\,{\sc ii}] emission for Galactic latitudes $b=0$\degr,
 $\pm$0.5\degr, and $\pm$1.0\degr. We combine these maps with those of
 H\,{\sc i}, $^{12}$CO, and $^{13}$CO to separate the different phases
 of the ISM and study their properties and distribution in the
 Galactic plane.  }
%
{ [C\,{\sc ii}] emission is mostly associated with spiral arms, mainly
emerging from Galactocentric distances between 4 and 10\,kpc.  It
traces the envelopes of evolved clouds as well as clouds that are in
the transition between atomic and molecular.  We estimate that most of
the observed [C\,{\sc ii}] emission is produced by dense photon
dominated regions ($\sim$47\%), with smaller contributions from
CO--dark H$_2$ gas ($\sim$28\%), cold atomic gas ($\sim$21\%), and
ionized gas ($\sim$4\%).  Atomic gas inside the Solar radius is mostly
in the form of cold neutral medium (CNM), while the warm neutral
medium (WNM) gas dominates the outer galaxy. The average fraction of
CNM relative to total atomic gas is $\sim$43\%.  We find that the warm
and diffuse CO--dark H$_2$ is distributed over a larger range of
Galactocentric distances (4--11\,kpc) than the cold and dense H$_2$
gas traced by $^{12}$CO and $^{13}$CO (4--8\,kpc). The fraction of
CO-dark H$_2$ to total H$_2$ increases with Galactocentric distance,
ranging from $\sim$20\% at 4\,kpc to $\sim$80\% at 10\,kpc. On
average, CO-dark H$_2$ accounts for $\sim$30\% of the molecular mass
of the Milky Way. When the CO--dark H$_2$ component is included, the
radial distribution of the CO--to--H$_2$ conversion factor is steeper
than that when only molecular gas traced by CO is considered.  Most of
the observed [C\,{\sc ii}] emission emerging from dense photon
dominated regions is associated with modest far--ultraviolet fields in
the range $\chi_0\simeq1-30$.

}
{} 

   \maketitle
%
%
%
%

\section{Introduction}
\label{sec:introduction}

The transition from atomic to molecular clouds, the formation of stars
within high density regions, the radiative feedback from newly formed
stars, and the disruption of molecular clouds and termination of star
formation are of great astrophysical interest as they are processes
governing the evolution of galaxies in our Universe.  In the Milky Way
the evolution of the interstellar medium (ISM) has traditionally been
studied with radio continuum and H$\alpha$ observations for the
ionized component \citep[e.g.][]{Haffner2009}, with observations of
the H\,{\sc i} 21\,cm line for the atomic gas component
\citep[e.g.][]{Kalberla2009}, and with rotational transitions of the
CO molecule for the dense molecular gas component \citep[e.g.][]{Dame01}.
But observations of the transition between these different stages of
evolution with velocity--resolved observations required to
separate different gas components along the line--of--sight (LOS) have been
missing.

The [C\,{\sc ii}] 158$\mu$m fine--structure transition is an excellent
tracer of the different stages of evolution of the ISM. 
As ionized carbon (C$^+$) can be found throughout the ISM and can be
excited by collisions with electrons, H\,{\sc i}, and H$_2$, the [C\,{\sc ii}]
line traces the warm ionized medium, the warm and cold diffuse atomic
medium, and warm and dense molecular gas. The [C\,{\sc ii}] 158$\mu$m
line is the main coolant of the diffuse ISM and therefore plays a key
role in the thermal instability that converts from  warm and
diffuse atomic clouds to cold and dense clouds. As the [C\,{\sc ii}]
intensity is sensitive to column density, volume density, and kinetic
temperature, it can be used to determine the physical conditions of
the line--emitting gas.

The first sensitive all sky survey of line emission in the wavelength
range between 100$\mu$m and 1\,cm was carried out with the
Far-Infrared Spectrometer (FIRAS) on COBE \citep{Bennett1994}. The
angular resolution of COBE was 7\degr\ and the spectral resolution of
FIRAS was about 1000 km\,s$^{-1}$ for the [C\,{\sc ii}] 158$\mu$m and
[N\,{\sc ii}] 205$\mu$m lines. The COBE map showed that [C\,{\sc ii}]
is widespread over the Galactic plane and that it is the brightest
far-infrared line. COBE also found a correlation between the [C\,{\sc
ii}], [N\,{\sc ii}], and far--infrared (FIR) continuum emission,
suggesting that [C\,{\sc ii}] might be a good tracer of star-formation
activity in galaxies. The balloon-borne BICE mission
\citep{Nakagawa1998} observed [C\,{\sc ii}] in the inner Galactic
plane with 15\arcmin\,angular resolution and 175 km s$^{-1}$ velocity
resolution. This mission gave better constraints on the latitudinal
distribution of the Galactic [C\,{\sc ii}] emission. The COBE and BICE missions
did not have the required angular and velocity resolution to resolve
spatially and in velocity individual clouds in the Galactic plane.
The Kuiper Airborne Observatory allowed the study of a handful of
H\,{\sc ii} regions with velocity-resolved [C\,{\sc ii}] observations
\cite[e.g][]{Boreiko1988,Boreiko91}.  These observations, and also
those recently obtained with {\it Herschel}
\citep[e.g.][]{Ossenkopf2012} and SOFIA \citep[e.g.][]{Graf2012}, were
mostly limited to hot and dense photon--dominated regions (PDRs)
associated with massive star-forming regions.

Far--infrared (FIR) dust continuum emission has been widely used as a
mass tracer of the ISM, and several high--resolution, large--scale,
sensitive maps of the Galactic plane exist
\citep{Carey2009,Molinari2010}.  Dust continuum emission is the result
of the integrated contribution of different ISM components distributed
along the line--of--sight. Therefore, it is impossible to use this
emission to isolate and study the different contributing ISM
components, particularly for observations of the Galactic plane.

 
The {\it Herschel} \citep{Pilbratt2010} Open Time Key Project Galactic Observations of
Terahertz C+ (GOT\,C+; \citealt{Langer2011}) provides  the first  high--angular
resolution (12\arcsec), velocity--resolved (0.1 km s$^{-1}$)
observations of the [C\,{\sc ii}] line in the Galactic disk, allowing us to resolve individual
clouds and to separate the different ISM components along the
line--of--sight. The GOT\,C+ survey consists of $\sim$500 LOSs
distributed in a volume-weighted sampling of the Galactic disk.  We
have complemented our GOT\,C+ [C\,{\sc ii}] data with
observations of the H\,{\sc i} 21 cm line, tracing diffuse atomic gas,
and of CO and its isotopologues, tracing the dense molecular gas. 

In this paper we present the full GOT\,C+ Galactic plane survey. We
use the kinematic information provided by high velocity resolution
observations of [C\,{\sc ii}], obtained  with the HIFI \citep{deGraauw2010}
instrument on {\it Herschel},  together with complementary CO and
H\,{\sc i} surveys to study the structure of the galaxy and to
characterize the distribution of the different components of the
interstellar medium in the Milky Way.

Neutral atomic hydrogen represents the dominant gas  component of the
interstellar medium in galaxies. This H\,{\sc i} gas is predicted to
exist in two phases in rough thermal pressure equilibrium
\citep{Pikelner1968,Field1969,Wolfire1995}; the cold neutral medium
($n_{\rm H} \simeq 50$\,cm$^{-3}$; $T_{\rm kin}\simeq$80\,K; CNM) and
the warm neutral medium ($n_{\rm H} \simeq 0.5 $\,cm$^{-3}$; $T_{\rm
kin}\simeq$8000\,K; WNM). Note, however, that gas in thermally
unstable conditions has been observed
\citep[e.g][]{Dickey1977,Heiles2003} and predicted in numerical
simulations \citep[see][and references therein]{Vazquez-Semadeni2012}.
The atomic gas component has been extensively mapped using the H\,{\sc
i} 21\,cm line in the Milky Way and external galaxies.  But this line
in emission traces only the total column of gas along the
line--of--sight (assuming that it is optically thin), and therefore it
is impossible to disentangle the relative contribution from gas in the
CNM or WNM.  H\,{\sc i} absorption studies of nearby clouds in the
foreground of extragalactic continuum sources have provided
constraints on the relative fraction of the CNM and WNM, suggesting
that 60\% of the gas is in the WNM and 40\% is in the CNM
\citep{Heiles2003}. But these observations are limited to a small
number of sources distributed mostly at high Galactic latitudes, thus
corresponding only to the local ISM.  We therefore lack the knowledge
about the distribution and conditions of the CNM and WNM over the
entire Galactic disk. The [C\,{\sc ii}] 158$\mu$m line is also a
tracer of the diffuse neutral gas but has the advantage that it is
also sensitive to the density and temperature of the gas. In
particular, because of the density contrast between the CNM and WNM,
for a given H\,{\sc i} column density the [C\,{\sc ii}] emission
associated with the WNM is expected to be a factor of $\sim$20 weaker
than that associated with the CNM  \citep{Wolfire2010b}. We can therefore use the GOT\,C+
[C\,{\sc ii}] emission in combination with H\,{\sc i} to study the
distribution and properties of the WNM and CNM over the entire
Galactic disk.

The mass of molecular clouds has traditionally been traced using
observations of the CO molecule.  However, the column density from the
cloud exterior to the location where hydrogen becomes molecular is
smaller than that required for carbon to become molecular in the form
of CO. It is therefore expected that in the outer parts of molecular
clouds a significant fraction of the H$_2$ is associated with C$^+$
and C$^0$ instead of with CO. This gas is called ``hidden--H$_2$'' or
``dark--H$_2$'' gas
\citep[see][]{Grenier2005,Langer2010,Wolfire2010}. In this paper we
will refer to this component as ``CO--dark H$_2$ gas".  Indirect
observations of gamma--rays \citep{Grenier2005} and dust continuum
emission \citep{Bernard2011} suggest that 30--50\% of the molecular
mass in the Galaxy is not traced by CO.  The preliminary results from
GOT\,C+, based on a sample of 16 LOSs, reached a similar conclusion
\citep{Langer2010,Velusamy2010}.  \citet{Velusamy2012} used a Gaussian
decomposition approach to study the GOT\,C+ LOS in the inner Galaxy,
$-90$\degr$<l<57$\degr, discussing the distribution of the [C\,{\sc
ii}] velocity components as a function of Galactocentric distance and
the distribution of CO--dark H$_2$ gas as function cloud type (diffuse
clouds, transition clouds, and dense clouds traced by $^{13}$CO).  A
more detailed analysis of the Gaussian decomposition of the GOT\,C+
[C\,{\sc ii}] emission in the inner galaxy will be presented in Langer
et al. (2013) in preparation.  Note that some of the H$_2$ gas might
not be traced by CO due to the lack of sensitivity of the CO
observations. In the Taurus molecular cloud, \citet{Pineda2010a} found
that 23\% of the total mass is in the outskirts of the cloud where
neither $^{12}$CO nor $^{13}$CO could be detected in single pixels,
but both isotopologues could be readily measured by averaging pixels
over a large area.

A significant fraction of the FIR line and continuum emission in
galaxies originates from dense photon--dominated regions (or
photodissociation regions; or PDRs; see \citealt{HollenbachTielens99}
and references therein). In PDRs the chemistry and thermal balance is
dominated by the influence of far--ultraviolet photons from massive
stars. Therefore, PDRs are the best sites for studying the radiative
feedback from newly formed massive stars on their progenitor molecular
gas, a process that plays a key role in the regulation of
star--formation in galaxies. The [C\,{\sc ii}] line is a very bright
tracer of PDRs and, together with fine--structure transitions of
[O\,{\sc i}] and [C\,{\sc i}] and rotational transitions of CO, can be
used to determine the physical conditions of the line--emitting
regions. The FUV field intensity is a key parameter for PDRs and is
closely related to the massive star--formation activity around
PDRs. The FUV field is often measured in units of the \citet{Draine78}
field\footnote{ The average FUV intensity of the local ISM is
2.2$\times10^{-4}$\,erg\,cm$^{-2}$\,s$^{-1}$\,sr$^{-1}$
\citep{Draine78}. The Draine field is isotropic (i.e. a given point is
illuminated from 4$\pi$ steradians), while the surface of the clouds
considered here are only illuminated from 2$\pi$ steradians, therefore
the rate of photoreactions at the cloud surface are half of what they
would be with the Draine field.} defined as the average FUV intensity
in the local ISM.  \citet{Kaufman99} showed that the ratio of [C\,{\sc
ii}] to CO is a good tracer of the strength of the FUV field in the
galaxy.  Using this line ratio in a sample of 16 LOSs,
\citet{Pineda2010b} found that most of the dense PDRs have FUV fields
lower than 100 times the interstellar FUV field. With the complete
GOT\,C+ survey, we can use the [C\,{\sc ii}]/CO ratio to trace the
strength of the FUV radiation field over the entire Galactic disk.

This paper is organized as follows. In Section~\ref{sec:observations}
we describe the GOT\,C+ observations and data reduction. We also
describe the observations of ancillary CO and H\,{\sc i} data. In
Section~\ref{sec:posit-veloc-maps} we present position--velocity maps
of the [C\,{\sc ii}] emission and relate them to the spiral structure
of the Milky Way.  In Section~\ref{sec:radi-distr-plots}, we present
the distribution of the [C\,{\sc ii}], H\,{\sc i}, and CO emissivities
as a function of Galactocentric distance.  In
Section~\ref{sec:origin-c-sc}, we combine the [C\,{\sc ii}], H\,{\sc
i}, and CO emission to identify the different ISM components and
determine their contribution to the observed [C\,{\sc ii}] emission.
We describe the distribution of the warm and cold neutral medium
atomic gas components in Section~\ref{sec:disent-cnm-wnm}. In
Section~\ref{sec:distr-h_2-plane}, we study the radial distribution of
the CO--dark H$_2$ gas component and compare it to that of molecular gas
traced by CO and $^{13}$CO. In Section~\ref{sec:fuv-radiation-field}
we present a determination of the FUV field distribution in the
Galactic plane.  We summarize our results in
Section~\ref{sec:conclusions}.

\begin{figure}
   \centering
   \includegraphics[width=0.5\textwidth]{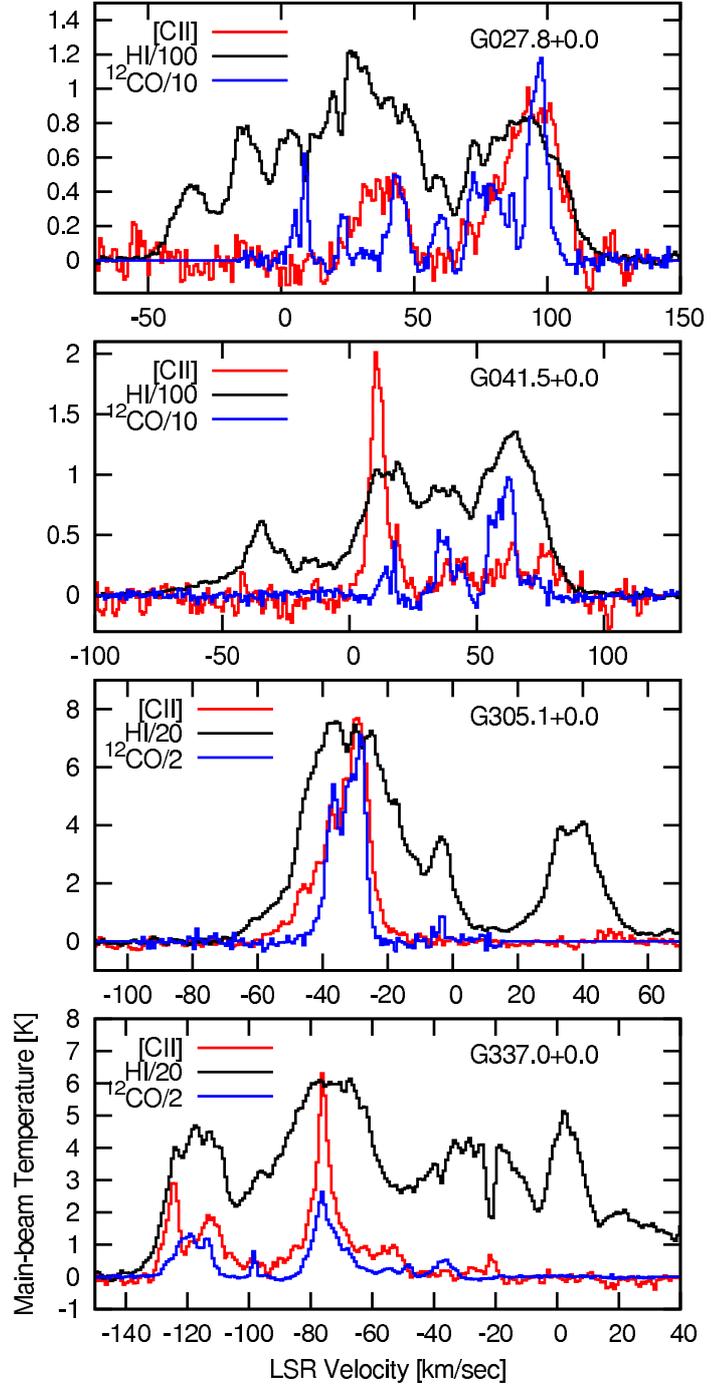}
 \caption{A selection of GOT\,C+ lines--of--sight observed in
   [C\,{\sc ii}], H\,{\sc i}, and $^{12}$CO emission. Note that the
   [C\,{\sc ii}] emission systematically extends closer to the tangent
   velocity compared with the $^{12}$CO line, where it follows the
   distribution of the H\,{\sc i} emission (see
   Section~\ref{sec:posit-veloc-maps}). }
    \label{fig:spectra}
   \end{figure}

  \begin{figure*}
   \centering
   \includegraphics[width=0.9\textwidth]{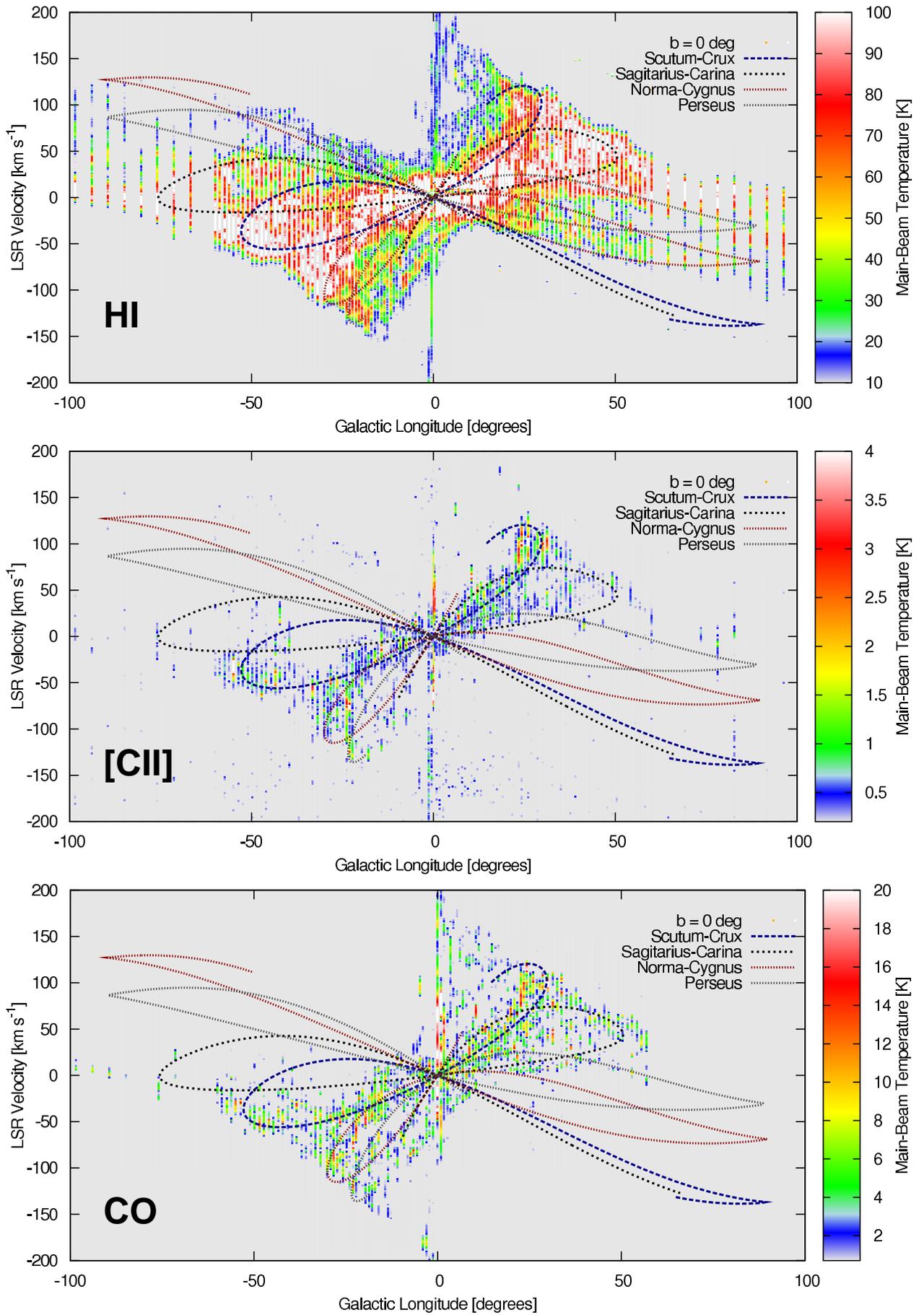}
      \caption{Position velocity maps of H\,{\sc i}, [C\,{\sc ii}],
              and $^{12}$CO $J = 1 \to 0$ for the observed GOT\,C+ LOSs
              between $l=-$100\degr\ and 100\degr\ at $b= 0$\degr. We have
              overlaid the Scutum-Crux, Sagittarius Carina, Perseus,
              and Norma-Cygnus spiral arms projected onto the
              position--velocity maps (see
              Section~\ref{sec:posit-veloc-maps}). }
\label{fig:pvmap_spirals}
   \end{figure*}
  \begin{figure*}
   \centering
   \includegraphics[width=\textwidth]{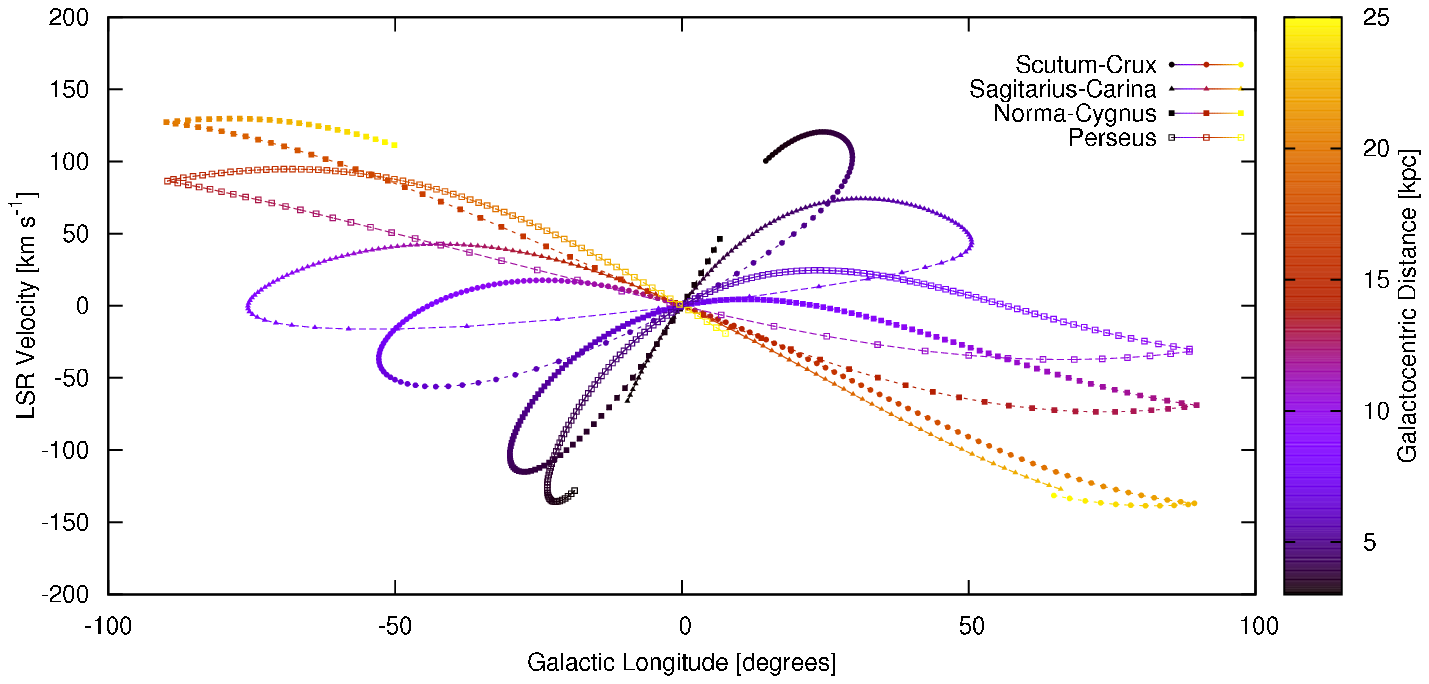}
      \caption{Projections in the Galactic longitude--velocity map of
              the Scutum-Crux, Sagittarius Carina, Perseus, and
              Norma-Cygnus spiral arms (see
              Section~\ref{sec:posit-veloc-maps}).  The distance to
              the Galactic center is color coded according to the
              scale shown at the right.
}
\label{fig:pvdist}
   \end{figure*}

 \section{Observations}
\label{sec:observations}

\subsection{[C\,{\sc ii}] observations}\label{sec:c-sc-ii}

We surveyed the Galactic plane in the [C\,{\sc ii}] $^2$P$_{3/2} \to
^2$P$_{1/2}$ fine structure line at 1900.5469\,GHz (rest frequency)
with the HIFI \citep{deGraauw2010} instrument on--board the {\it
Herschel} space observatory \citep{Pilbratt2010}. These observations
are part of the {\it Herschel} Open Time Key Project Galactic
Observations of Terahertz C+ (GOT\,C+), which is devoted to studying
the [C\,{\sc ii}] 158\,$\mu$m line over a wide range of environments
in the Galactic plane, the Galactic center, and in the Outer Galaxy
Warp. The GOT\,C+ Galactic plane survey consists of 452 LOSs
distributed in an approximately volume--weighted sample of the
Galactic plane. We present a selection of GOT\,C+ lines--of--sight in
[C\,{\sc ii}], H\,{\sc i}, and $^{12}$CO emission in
Figure~\ref{fig:spectra}.  The entire data set is presented as a set
of position--velocity maps in Figure~\ref{fig:pvmap_spirals} and in
Appendix~\ref{sec:posit-veloc-maps-4}.  We sampled in Galactic
longitude every 0.87\degr\ for $|l|<60$\degr, every 1.3\degr\ for
$30$\degr$<|l|<60$\degr, and every 4.5\degr\ for
$60$\degr$<|l|<90$\degr.  In the Outer Galaxy ($|l|>90$\degr), we
sampled in alternating steps varying from 4.5\degr\ to 13.5\degr.  We
sampled in Galactic latitude 0.5\degr\ and 1.0\degr\ above and below
the plane at alternate positions in $l$. In 11 LOSs in the Outer
Galaxy with $|l|>90$\degr, we sampled $b=\pm2.0$\degr\ instead of
$\pm0.5$\degr.

The [C\,{\sc ii}] 1.9\,THz observations were carried out with the HIFI
Band 7b receiver, which is employs Hot Electron Bolometer (HEB)
mixers.  The HEB bands in HIFI show prominent electrical standing
waves that are produced between the HEB mixing element and the first
low noise amplifier. The standing wave shape is not a standard
sinusoid and is difficult to remove from the resulting spectrum using
standard fitting methods \citep{Higgins2009}. To remove these standing
waves we used a procedure developed at the HIFI Instrument Control
Center (ICC; Ian Avruch priv. comm.) which generates a library of
standing wave shapes from different observations and finds the best
fitting one to correct the observed spectrum (see
\citealt{Higgins2011} for a detailed description of this method). We
processed our data with this standing wave removal script in {\tt
HIPE} \citep{Ott2006} version 8. In {\tt HIPE} we also removed
residual standing waves by fitting a single sinusoidal function using
the {\tt FitHIFIFringe()} procedure. After all standing waves are
removed, we exported our data to the {\tt CLASS90}\footnote{{\tt
http://www.iram.fr/IRAMFR/GILDAS}} data analysis software, which we
used to combine different polarizations, fit polynomial baselines
(typically of order 3), and smooth the data in velocity.

The angular resolution of the [C\,{\sc ii}] observations is
12\arcsec. We applied a main--beam efficiency of 0.72 to transform the
data from an antenna temperature to a main-beam temperature scale
\citep{Roelfsema2012}.

We used the data produced by the wide band spectrometer (WBS), which
have a channel width of 12\,MHz (0.16 km s$^{-1}$ at 1.9 THz).  We
later smoothed the data to a channel width of 0.8 km s$^{-1}$.  For
this channel width the average rms noise of our data is\footnote{
For the typical [C\,{\sc ii}] FWHM line width of about 3\,km s$^{-1}$
(Langer et al. 2013 in preparation), this sensitivity limit
corresponds to $1.1\times10^{-6}$ erg s$^{-1}$ cm$^{-2}$
sr$^{-1}$. The integrated intensity in units of K\,km\,s$^{-1}$ can be
converted that in units of erg s$^{-1}$ cm$^{-2}$ sr$^{-1}$ using
$I$[K\,km\,s$^{-1}$]=$1.43\times10^{5}$$I$[erg s$^{-1}$ cm$^{-2}$
sr$^{-1}$] \citep{Goldsmith2012}. } 0.1\,K.

As we expected the [C\,{\sc ii}] emission to be extended in the
Galactic plane, we used the LoadChop with reference observing mode.
In this observing mode an internal cold calibration source is used as
a comparison load to correct short term changes in the instrument
behavior.  Since the optical path differs between source and internal
reference, a residual standing wave structure remains and is corrected
with an observation of a sky reference position.  For a given Galactic
longitude, LOSs with $b=0$\degr, $\pm0.5$\degr\, and $\pm1.0$\degr\
share the same reference position at $b=\pm2.0$\degr. In the 11 Outer
Galaxy positions where we observed $b=0$\degr, $\pm1.0$\degr, and
$\pm2.0$\degr, we used the $b=\pm2.0$\degr\ as a reference for
$b=0$\degr, $\pm1.0$\degr\ and $b=\pm4.0$\degr\ as a reference for
$b=\pm2.0$\degr. We recovered the reference spectrum (OFF) using {\tt
HIPE} by subtracting from the ON--OFF spectrum a spectrum for which the
reference subtraction step in the HIFI pipeline was suppressed. The
resulting spectra include a standing wave with period $\sim$90\,MHz.
We combined all observations of the reference spectrum in {\tt CLASS90
} and removed the standing wave using a Fast Fourier Transform (FFT)
procedure. In the cases where emission in the reference position is
present, we decomposed this emission into Gaussian components and
corrected the ON--source spectra appropriately. A total of 51\% of the
GOT\,C+ spectra were corrected for emission in the reference position.


\subsection{CO observations}\label{sec:mopra-observations}

To complement the GOT\,C+ data, we observed the $J = 1 \to 0$
transitions of $^{12}$CO, $^{13}$CO, and C$^{18}$O with the ATNF
Mopra\footnote{The Mopra radio telescope is part of the Australia
Telescope which is funded by the Commonwealth of Australia for
operation as a National Facility managed by CSIRO.}  Telescope. We
observed all GOT\,C+ positions towards the inner Galaxy between
$l=-175.5$\degr and $l=56.8$\degr\ with an angular resolution of
33\arcsec.  Typical system temperatures were 600, 300, and 250\,K for
$^{12}$CO, $^{13}$CO, and C$^{18}$O, respectively.  To convert from
antenna to main--beam temperature scale, we used a main-beam
efficiency of 0.42 \citep{Ladd05}. All lines were observed
simultaneously with the MOPS spectrometer in zoom mode. The spectra
were smoothed in velocity to 0.8\,km s$^{-1}$ for $^{12}$CO and
$^{13}$CO and to 1.6\,km\,s$^{-1}$ for C$^{18}$O. The typical rms
noise is 0.6\,K for $^{12}$CO and 0.1\,K for both $^{13}$CO and
C$^{18}$O.  We checked pointing accuracy every 60 minutes using the
closest and brightest SiO maser. For $102$\degr$ < l < 141$\degr, we
used the $^{12}$CO $J=1\to0$ data from the FCRAO Outer Galaxy survey
\citep{Heyer1998a}. The data have an angular resolution of 45\arcsec,
and a rms noise of about 0.6\,K in a 0.8\,km\,s$^{-1}$ channel.

\subsection{H\,{\sc i} data }

We used supplementary H\,{\sc i} 21\,cm data taken from the Southern
Galactic Plane Survey \citep[SGPS;][]{McClure-Griffiths2005}, the
Canadian Galactic Plane Survey \citep[CGPS;][]{Taylor2003}, and the
VLA Galactic Plane Survey \citep[VGPS;][]{Stil2006}. These surveys
combined cover the 253\degr$<l< 67$\degr\ portion of the Galactic
plane. The SGPS (253\degr$<l<$358\degr) data have an angular
resolution of 2\arcmin, and a rms noise of 1.6\,K per 0.8 km s$^{-1}$
channel. The CGPS covers 74\degr$<l<$147\degr, with an angular
resolution of 1\arcmin, and a rms noise of 2\,K in a
0.82\,km\,s$^{-1}$ channel. Finally, the coverage of the VGPS is
18\degr$<l<$67\degr, with an angular resolution of 1\arcmin, and a rms
noise  of 2\,K in a 0.82 km s$^{-1}$ channel.  For the Galactic
center we used the data presented by \citet{McClure-Griffiths2012}
which have parameters similar to those of the SGPS. We also used the
GALFA--H\,{\sc i} Arecibo survey \citep {Peek2011} to cover the
Galactic longitude range between 180\degr$<l<$212\degr. The
GALFA--H\,{\sc i} data have an angular resolution of 4\arcmin\ and a
typical rms noise of 80 mK in a 1\,km\,s$^{-1}$ channel.



\section{The Distribution of [C\,{\sc ii}] in the Milky Way}
\label{sec:distribution-c-}

\subsection{Position--Velocity Maps}
\label{sec:posit-veloc-maps}

In Figure~\ref{fig:pvmap_spirals} we present the GOT\,C+ survey in a
set of position-velocity maps for $b=$ 0\degr.  The corresponding
position--velocity maps for $\pm$0.5\degr, and $\pm$1.0\degr\ are
presented in Appendix~\ref{sec:posit-veloc-maps-4}. The observed
[C\,{\sc ii}] emission peaks at $b=$ 0\degr\ and decreases for LOSs
above and below the Galactic plane. Most of the [C\,{\sc ii}] is
concentrated in the inner galaxy ($-60$\degr$\leq l \leq$60\degr).
The position--velocity map is overlaid with projections of the
Scutum-Crux, Sagittarius-Carina, Perseus, and
Norma-Cygnus\footnote{The Norma-Cygnus arm is also known as the
Norma-3\,kpc arm \citep[e.g.][]{Vallee2008}. } Milky Way spiral
arms. We used the fits to the parameters determining the logarithmic
spiral arms presented by \citet{Steinman-Cameron2010} and assumed a
flat rotation curve the distance of the Sun to the Galactic center,
$R_{\odot}$ = 8.5\,kpc, and an orbital velocity of the sun with
respect to the Galactic center, $V_{\odot}$ = 220\,km\,s$^{-1}$, which
are the values recommended by the International Astronomical Union
(IAU). We show the projected spiral arms in the range of
Galactocentric distance between 3\,kpc and 25\,kpc. {\it We can see
that most of the [C\,{\sc ii}] emission is closely associated with the
spiral arms and it is brightest at the spiral arms' tangent
points}. In Figure~\ref{fig:pvdist}, we plot the spiral arms with a
color scale that represents the Galactocentric distance. Comparing
Figure~\ref{fig:pvdist} with the panels in
Figure~\ref{fig:pvmap_spirals}, we see that the [C\,{\sc ii}] emission
is mostly concentrated within the inner 10\,kpc of the Galaxy (see
also Section~\ref{sec:radi-distr-plots}).  The [C\,{\sc ii}] emission
is slightly shifted with respect to that of $^{12}$CO, with the
[C\,{\sc ii}] emission present at velocities that are closer to the
tangent velocity.  This shift can also be seen near the tangent
velocities in the sample spectra shown in Figure~\ref{fig:spectra}. 
This suggests that the [C\,{\sc ii}] emission traces an intermediate
phase between the extended H\,{\sc i} emission and the cold and dense
molecular gas traced by $^{12}$CO.

   \begin{figure}
   \centering

   \includegraphics[width=0.5\textwidth]{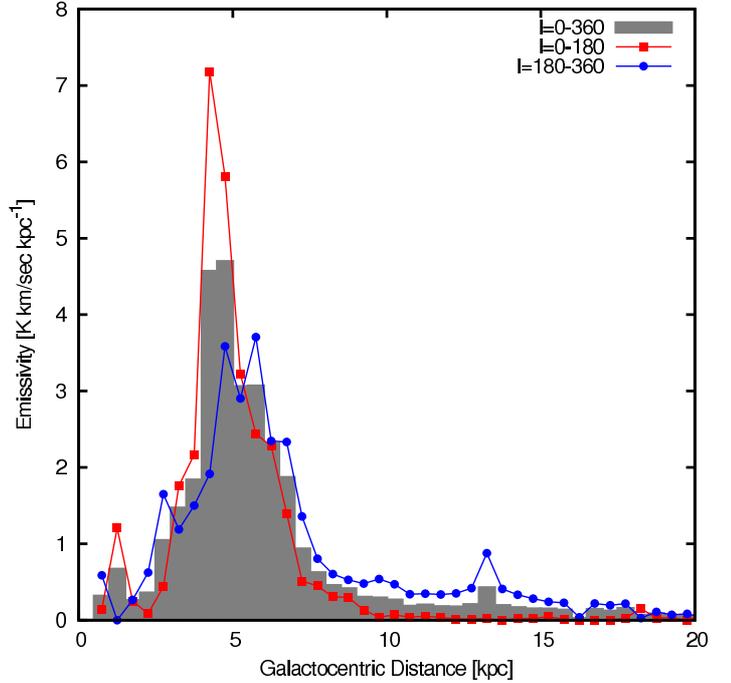}
  \caption{Radial distribution of the azimuthally averaged [C\,{\sc
  ii}] emissivity at $b=0$\degr. The shaded area represents the
  emissivity calculated considering all sampled Galactic
  longitudes. The connected boxes represent the radial emissivity
  distribution for $0$\degr$<l<180$\degr, while the connected circles
  represent that for $180$\degr$<l<360$\degr. The typical 1$\sigma$
  uncertainty in the [C\,{\sc ii}] emissivity is
  0.02\,K\,km\,s$^{-1}$\,kpc$^{-1}$. }
         \label{fig:radial_distro}
   \end{figure}

   \begin{figure}
   \centering
   \includegraphics[width=0.5\textwidth]{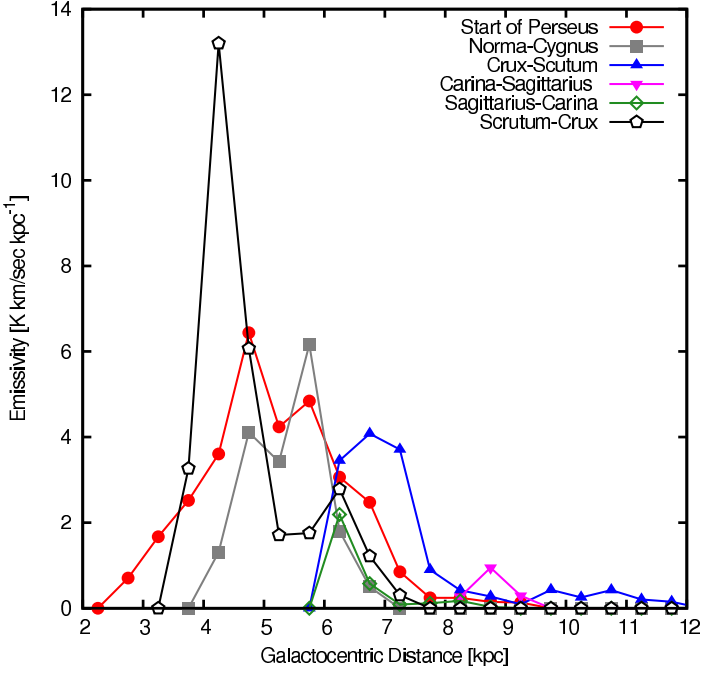}
 \caption{Radial distribution of the azimuthally averaged [C\,{\sc
  ii}] emissivity for GOT\,C+ LOSs toward the tangents of spiral arms
  as defined by \citet{Vallee2008}. }
         \label{fig:tangents}
   \end{figure}

 \begin{figure}
   \centering
   \includegraphics[width=0.5\textwidth]{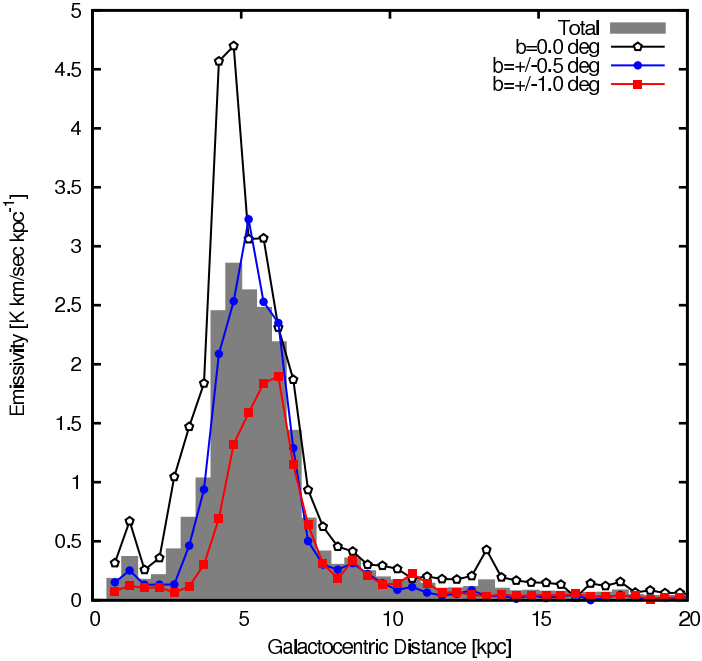}
  \caption{Radial distribution of the azimuthally averaged [C\,{\sc
  ii}] emissivity for LOSs at $b=0$\degr, $b=\pm0.5$\degr,
  $b=\pm1.0$\degr, and for all observed LOSs together.}
\label{fig:radial_distro_pm05_pm1.0}
   \end{figure}

\subsection{Radial Distribution Plots}
\label{sec:radi-distr-plots}

In the following we study the distribution of the [C\,{\sc ii}] emission as a
function of Galactocentric radius.  We divided the galaxy in a set of
rings with a distance to the Galactic center $R_{\rm gal}$ and width
$\Delta R=0.5$\,kpc. For each of these rings, we calculated the
azimuthally--averaged integrated intensity per unit length or
emissivity of the [C\,{\sc ii}] line. We describe the procedure used
to determine these emissivities in
Appendix~\ref{sec:determ-azim-aver}.

In Figure~\ref{fig:radial_distro} we present the azimuthally averaged
emissivity of [C\,{\sc ii}] as a function of Galactocentric radius.
The [C\,{\sc ii}] emission shows a main peak at 4.2 kpc and a
secondary peak at $\sim6$ kpc. As mentioned above, the [C\,{\sc ii}]
emission is concentrated in the inner 10\,kpc of the Galaxy. When we
separate the contribution from LOSs with $l=0-180$\degr\ and
$l=180-360$\degr, we can see that the peak at 4.2 kpc originates from
LOSs in the $l=0-180$\degr\,range while the peak at 6 kpc originates
from LOSs in the $l=180-360$\degr\ range.  This difference in the
emissivity distribution between both sides of the Galaxy illustrates
the magnitude of the intrinsic scatter within the rings, which is a
result of the fact that the Galaxy is not axisymmetric.  Because the
aim of our analysis is to describe the average properties of the ISM
within a Galactocentric ring, we only consider the uncertainties that
describe how accurately this azimuthal average emissivity is
determined. These uncertainties are generally very small, as they are
the result of averaging over a large number of data values. In many
figures the error--bars are too small to be visible, especially when
the emissivity vary over a large range, and we will quote them in the
figure captions. Note that in most cases the uncertainties in the
assumptions of our modelling are much larger than those indicated by
the observational error--bars.

 In Figure~\ref{fig:tangents} we show the contributions to the
[C\,{\sc ii}] emissivity from the different spiral arm tangents. We
consider LOSs in 12\degr--wide bins in Galactic longitude centered at
the spiral arm tangent locations defined by \citet{Vallee2008}. Note
that different spiral arms can be seen in this set of LOSs, but the
peak of the [C\,{\sc ii}] emission corresponds to the tangent of the
spiral arm.  The peak at 4.2\,kpc is associated with
the Scutum--Crux tangent while the 6\,kpc peak is related to the
combined emission from the Start of Perseus (peak at 4.7 kpc),
Norma-Cygnus (peak at 5.7 kpc), and Crux--Scutum (peak at 6.7 kpc)
tangencies. The Sagittarius--Carina (peak at 6.2 kpc) and
Carina--Sagittarius (peak at 8.7\,kpc) make a smaller
contribution. Figure~\ref{fig:tangents} shows that the brightest
[C\,{\sc ii}] emission in the GOT\,C+ survey arises from the
Scutum--Crux tangent at 4.2 kpc.

In Figure~\ref{fig:radial_distro_pm05_pm1.0} we show the radial
distribution of the [C\,{\sc ii}] emissivity from LOSs with
$b=0$\degr, $b=\pm 0.5$\degr, and $b=\pm 1.0$\degr, and for all
observed LOSs together. The [C\,{\sc ii}] peak at 4.2\,kpc from the
Scutum-Crux arm is only seen for $b=0$\degr, while the intensities
from $b=\pm0.5$\degr, $b=\pm 1.0$\degr\ mostly contribute to the
intensity at 5.2\,kpc from the Galactic center. The shift in the peak
of the radial distribution for $|b|>$0\degr is related to the vertical
structure of the Galactic disk. A cloud with $R_{\rm gal}=5$\,kpc
observed in a LOS with $|b|$=1\degr\ would have a vertical distance
from the Galactic plane of $\sim$65\,pc, assuming that it is in the
near side of the Galaxy, while a cloud with $R_{\rm gal}=4$\,kpc, in
the same situation, would have a vertical distance of $\sim$100\,pc. A
Gaussian fit to the latitudinal [C\,{\sc ii}] distribution observed by
BICE \citep[see Fig. 9 in][]{Nakagawa1998} results in an angular FWHM
of the Galactic disk of 1.9\degr. Assuming that most of their observed
[C\,{\sc ii}] emission is in the near side of the Galaxy at a
heliocentric distance of about 4\,kpc, this angular FWHM corresponds
to 130\,pc. Our inability to detect a peak of emission at 4\,kpc for
$|b|>0$\degr\ seems to be consistent with this value of the [C\,{\sc
ii}] disk thickness. Results on the vertical structure of the Galactic
disk using the GOT\,C+ data will be presented in a future paper.

\begin{figure}
   \centering
   \includegraphics[width=0.5\textwidth]{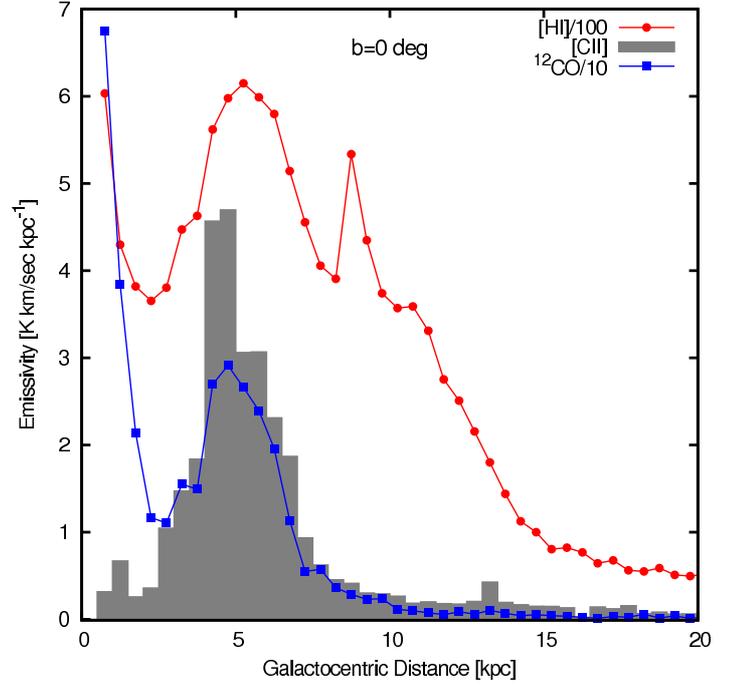}
  \caption{Radial distribution of the azimuthally averaged [C\,{\sc
  ii}], H\,{\sc i}, and $^{12}$CO emissivities at $b=0$\degr. The
  discontinuity in the H\,{\sc i} emissivity distribution at
  $\sim$9\,kpc is discussed in
  Section~\ref{sec:disent-cnm-wnm}. Typical 1$\sigma$ uncertainties
  in the [C\,{\sc ii}], H\,{\sc i}, and $^{12}$CO emissivities (before scaling) are
  0.02, 0.6, and 0.1 K\,km\,s$^{-1}$\,kpc$^{-1}$, respectively. }
         \label{fig:hi_co_cii}
   \end{figure}

In Figure~\ref{fig:hi_co_cii}, we compare the radial distributions of
the [C\,{\sc ii}], $^{12}$CO, and H\,{\sc i} emissivities. Typical
1$\sigma$ uncertainties in the [C\,{\sc ii}], H\,{\sc i}, and
$^{12}$CO emissivities are 0.02, 0.6, and 0.1
K\,km\,s$^{-1}$\,kpc$^{-1}$, respectively. We see that both [C\,{\sc ii}] and
$^{12}$CO are mostly concentrated in the 4--10\,kpc range, peaking at
4.25\,kpc. The distribution of the H\,{\sc i} emissivities extends
over a larger range of Galactocentric distances, with a peak at
5\,kpc.

  \begin{figure*}
   \centering
   \includegraphics[width=\textwidth]{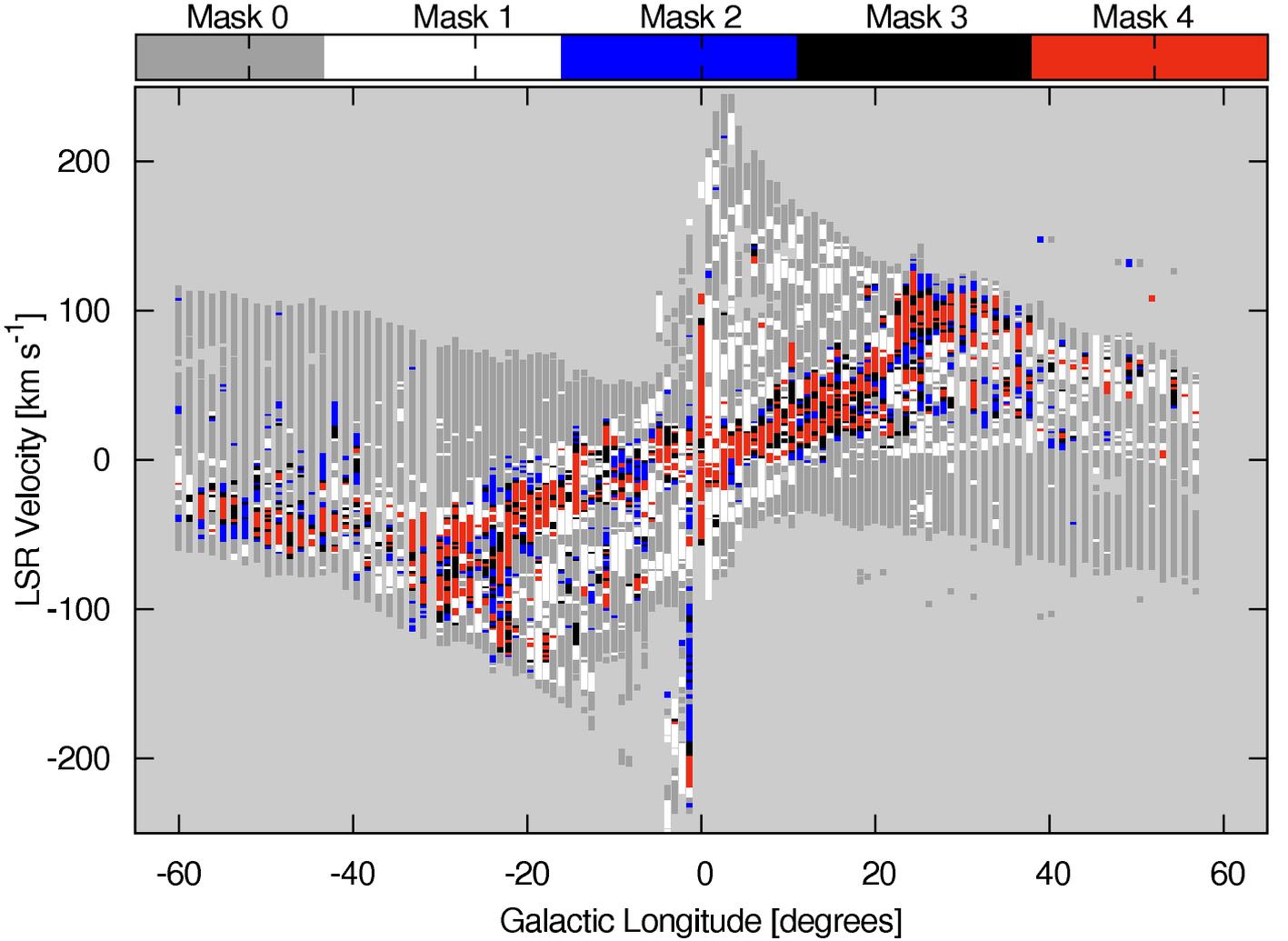}
      \caption{Position velocity maps of the different types of
              regions defined in Section~\ref{sec:velocity-components}
              at $b=0$\degr. Mask 0 ({\it grey}) represents
              velocity components with only H\,{\sc i} detected, Mask
              1 ({\it white}) are components with only H\,{\sc i} and
              CO detected, Mask 2 ({\it blue}) components with only
              H\,{\sc i} and [C\,{\sc ii}], Mask 3 ({\it black})
              components with H\,{\sc i}, [C\,{\sc ii}], and CO, and
              Mask 4 ({\it red}) components with H\,{\sc i}, [C\,{\sc
              ii}], $^{12}$CO, and $^{13}$CO.}
\label{fig:masks}
   \end{figure*}

\begin{table} [h]                                           
\caption{Definition of Mask Regions}
\label{tab:masks}
\centering                                                  
\begin{tabular}{c c c c c}                                 
\hline\hline
Mask & H\,{\sc i} & [C\,{\sc ii}] & $^{12}$CO & $^{13}$CO \\
\hline 
 0 & \checkmark & $\times$ & $\times$ & $\times$ \\
 1 & \checkmark & $\times$ & \checkmark & \checkmark$\times$ \\
 2 & \checkmark & \checkmark & $\times$ & $\times$ \\
 3 & \checkmark & \checkmark & \checkmark & $\times$  \\
 4 & \checkmark & \checkmark & \checkmark & \checkmark \\
\hline
\multicolumn{5}{l}{\checkmark: Emission is detected in individual spaxels.}\\
\multicolumn{5}{l}{$\times$: Emission is not detected.}\\
\multicolumn{5}{l}{\checkmark$\times$: Emission is either detected or not detected.}\\
\hline
\end{tabular}                                               
\end{table}  


   \begin{figure}[h]
   \centering
   \includegraphics[width=0.44\textwidth]{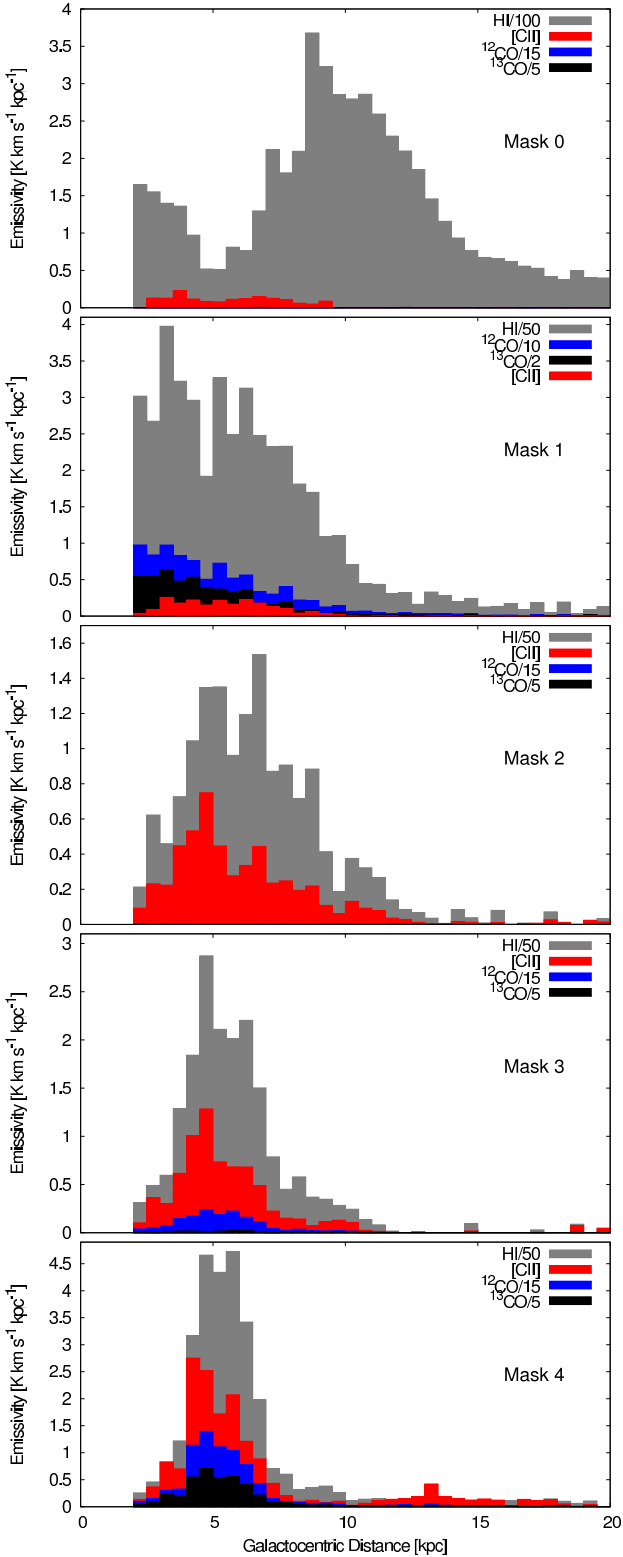}
      \caption{Galactocentric distribution of the different
              types of regions defined in
              Section~\ref{sec:velocity-components}. The plots include
              data from all LOSs at $b=0$\degr, $b=\pm 0.5$\degr, and
              $b=\pm 1.0$\degr.  }
         \label{fig:mask_radial_distro}
   \end{figure}
%


\section{The origin of the [C\,{\sc ii}] emission in the Galaxy}
\label{sec:origin-c-sc}

\subsection{Velocity Components}
\label{sec:velocity-components}
In the following, we associate the [C\,{\sc ii}] emission with other
tracers in order to separate the contribution from different ISM
phases to the observed emission. We define 5 mask regions that
represent different stages of the ISM evolution depending whether
H\,{\sc i}, [C\,{\sc ii}], $^{12}$CO, and $^{13}$CO are detected in an
individual pixel in the position--velocity map (spaxel).  We define
Mask\,0 as those spaxels in which H\,{\sc i} emission is detected but
[C\,{\sc ii}], CO, and $^{13}$CO not. These regions likely represent
low--volume density atomic gas (see
Section~\ref{sec:disent-cnm-wnm}). Mask\,1 includes spaxels in which
H\,{\sc i} and $^{12}$CO (or $^{13}$CO) are detected but [C\,{\sc ii}]
is not. These spaxels trace dense molecular gas that is too cold for
[C\,{\sc ii}] to be detectable at the sensitivity limit of the
observations. Mask\,2 are spaxels where H\,{\sc i} and [C\,{\sc ii}]
are detected but $^{12}$CO is not. These regions are likely larger
volume density atomic clouds that have insufficient FUV--photon
shielding to allow the build--up of CO and therefore in which most of
the gas--phase carbon is in the form of C$^+$. Mask\,3 includes
spaxels in which H\,{\sc i}, [C\,{\sc ii}], and $^{12}$CO are detected
but $^{13}$CO is not. These regions represent more shielded regions
where CO is forming but with inadequate column density to have
detectable $^{13}$CO emission at the sensitivity of our
observations. Finally, Mask\,4 includes spaxels in which H\,{\sc i},
[C\,{\sc ii}], $^{12}$CO, and $^{13}$CO are all detected. These regions
likely represent PDRs that have large volume and column densities and
are warm enough to produce significant [C\,{\sc ii}] emission. In
Table~\ref{tab:masks} we present a summary of the Mask region
definitions.

In Figure~\ref{fig:masks} we present the distribution of the different
Masks in the position--velocity map of the Galaxy for
$-$60\degr$<l<$60\degr\ and $b=0$\degr. We present similar
position--velocity maps for $b=\pm0.5$\degr\ and $\pm$1.0\degr\ in
Appendix~\ref{sec:posit-veloc-maps-4} (Figure~\ref{fig:mask2}).  The
dense PDRs (Mask\,4) are closely associated with the spiral arms. These
regions are surrounded with spaxels in Mask\,3 which, in turn, are
surrounded with spaxels in Mask\,2. We suggest that this spatial
arrangement indicates that Mask\,3 and Mask\,2 represent the envelopes
of the active star--forming regions traced by Mask\,4. There is a
significant fraction of spaxels in Mask\,1, particularly in the
molecular ring and local arms. Mask\,0 represents the largest fraction
of spaxels.

Figure~\ref{fig:mask_radial_distro} presents the intensity
distribution of H\,{\sc i}, [C\,{\sc ii}], $^{12}$CO, and $^{13}$CO
for the different Mask regions defined above for $b=0$\degr. Although
by definition Masks\,0 and 1 have no [C\,{\sc ii}] emission in a given
spaxel, its emission can still be seen when the data is averaged over
the different rings. The same result happens with $^{13}$CO in
Mask\,3.  The detection of emission after averaging the data in
azimuth shows that our classification into Masks depends on the
sensitivity of our observations. We will study the [C\,{\sc ii}]
emission in Mask 0 and 1 and the $^{13}$CO emission in Mask\,3
separately as they represent a different population of clouds that are
weak in [C\,{\sc ii}] and $^{13}$CO, respectively.

\subsection{Atomic gas}
\label{sec:atomic-gas}

We first assumed that all the observed [C\,{\sc ii}] emission arises
from a purely atomic gas at a kinetic temperature $T_{\rm
kin}=100$\,K.  For optically thin emission, the [C\,{\sc ii}]
intensity (in units of K\,km\,s$^{-1}$) is related to the C$^+$ column
density, $N_{C^+}$ ( cm$^{-2}$), and volume density of the collisional partner, $n$
(e$^-$, H, or H$_2$;  cm$^{-3}$), as \citep[see e.g.][]{Goldsmith2012},

\begin{equation}
\label{eq:3}
I_{\rm [CII] } = N_{{\rm C}^+} \left [3.05\times10^{15} \left (1+0.5
\left (1+\frac{A_{ul}}{R_{ul}n} \right )e^{91.21/T_{\rm kin}} \right
)\right ]^{-1},
\end{equation}
where $A_{\rm ul}=2.3\times10^{-6}$\,s$^{-1}$ is the Einstein
spontaneous decay rate and $R_{ul}$ is the collisional de--excitation
rate coefficient at a kinetic temperature $T_{\rm kin}$. We used a
value of $R_{ul}$ for collisions with H at $T_{\rm kin}=100$\,K of
$8.1\times10^{-10}$ s$^{-1}$\,cm$^{-3}$ \citep{Launay1977}.  In the
optically thin limit we can estimate the atomic gas column density,
$N({\rm H})$, from the H\,{\sc i} 21\,cm observations using, $N({\rm
H})=1.82\times10^{18} I({\rm HI})$\,cm$^{-2}$, with $I({\rm HI})$ in
units of ${\rm K\,km\,s^{-1}}$.  Observations of H\,{\sc i} absorption
toward continuum sources, however, suggest that the opacity of H\,{\sc
i} increases toward the inner Galaxy, where the assumption of
optically thin emission might not apply. \citet{Kolpak2002} derived an
average value, assuming $T_{\rm kin}=50$\,K, of $\tau\simeq1$ for
$R_{\rm gal}<8.5$\,kpc. For our assumed $T_{\rm kin}=100$\,K, this
opacity corresponds to $\tau\simeq0.5$.  We used this opacity to
correct the H\,{\sc i} column densities derived from the 21\,cm line
that is associated with [C\,{\sc ii}] emission, resulting in a 30\%
increase in the derived column density.

 The column density of ionized carbon can be estimated from
$N$(H\,{\sc i}) assuming an appropriate [C]/[H] fractional abundance,
provided that all gas--phase carbon is in the form of C$^{+}$.  There
is evidence that the gas--phase abundance of metals in the Milky Way
decreases with Galactocentric distance. \citet{Rolleston2000} found
that the fractional abundance distribution of several light elements
(C, O, Mg \& Si) can be represented by a linear function with slope
$-0.07\pm0.01$\,dex kpc$^{-1}$ over $6$\,kpc\,$<R_{\rm gal}<18$\,kpc .
We use this relationship to convert $N({\rm H})$ to $N({\rm C}^{+})$
for any given galactocentic distance, assuming that the slope of the
relative abundance gradient does not change significantly for $R_{\rm
gal}<6$\,kpc.  We assume a [C]/[H] fractional abundance of
$1.4\times10^{-4}$ for $R_{\rm gal}=8.5$\,kpc \citep{Cardelli1996}. The
[C]/[H] abundance as a function of Galactocentric distance is thus
given by
\begin{equation}
\label{eq:2}
[{\rm C}]/[{\rm H}]=5.5\times10^{-4}10^{-0.07R_{\rm gal}}.
\end{equation}
The [C]/[H] abundance drops by a
factor of 2.6 from $R_{\rm gal}=4$ kpc to 10\,kpc.

   \begin{figure}[t]
   \centering
   \includegraphics[width=0.5\textwidth]{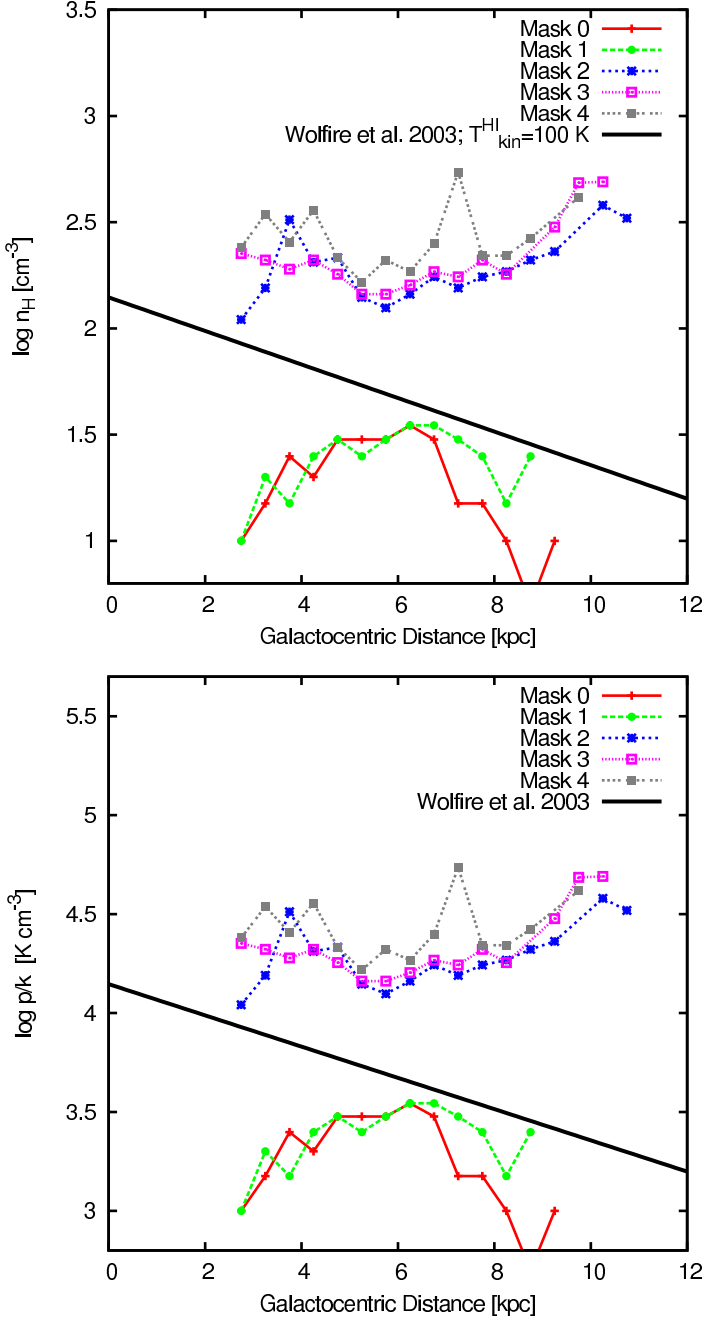}
      \caption{({\it Upper panel}) Neutral hydrogen volume density as
              a function of Galactocentric distance for the Mask
              regions defined in Section~\ref{sec:velocity-components},
              assuming that all [C\,{\sc ii}] is associated with
              atomic gas at $T_{\rm kin}=100$\,K. ({\it Lower panel})
              Corresponding thermal pressure as a function of
              Galactocentric distance.}
\label{fig:denstiy-pressure}
   \end{figure}

With the estimated value of $N(C^+)$ we used Equation~(\ref{eq:3}) to
solve for the volume density of H\,{\sc i} that is required to match
the observed [C\,{\sc ii}] emission, under the assumption that all
observed [C\,{\sc ii}] emission comes from atomic gas. In
Figure~\ref{fig:denstiy-pressure}a, we present the distribution of
hydrogen volume densities as a function of Galactocentric distance for
the different Mask regions for $4$\,kpc\,$<R_{\rm gal}<11$\,kpc, which
is the range of $R_{\rm gal}$ in which most of the [C\,{\sc ii}]
emission is detected.  The H volume density increases toward the
Galactic center with a range between 100\,cm$^{-3}$ to $10^{3}$
cm$^{-3}$ for Masks 2, 3 and 4. For Masks 0 and 1 the volume densities
are more moderate, ranging from 20 to 80\,cm$^{-3}$.
In Figure~\ref{fig:denstiy-pressure}b, we present the corresponding
thermal pressures ($p/k=nT_{\rm kin}$) as a function of Galactocentric
distance for the different Mask regions.  
We also show the thermal pressure distribution suggested by
\citet{Wolfire2003},
\begin{equation}
\label{eq:4}
p/k=1.4\times10^4\exp \left(\frac{-R_{\rm gal}}{5.5{\,\rm kpc}} \right ) {\rm K}\,{ \rm cm}^{-3},
\end{equation}
corresponding to the predicted thermal pressure of the CNM gas.  Masks
0 and 1 have thermal pressures that are consistent with those
suggested by \citet{Wolfire2003}. Thus, the densities and thermal
pressures estimated in Masks 0 and 1 are consistent with their [C\,{\sc
ii}] emission originating from the CNM.  Masks 2, 3, and 4 have volume
densities and thermal pressures that are about one order of magnitude
larger than those observed and predicted for the CNM.  We therefore
conclude that the assumption that the [C\,{\sc ii}] emission in Masks
2, 3, and 4 originates entirely from atomic gas is incompatible with
our observations.  As we will discuss below, part of the observed
[C\,{\sc ii}] emission in Masks 2, 3, and 4 also emerges from
molecular, and to a lesser extent, from ionized gas.

Motivated by the agreement between the pressures derived in Masks 0 and
1 and those predicted in Equation~(\ref{eq:4}), we used this
expression to estimate, assuming a constant $T_{\rm kin}=100$\,K, the
radial H volume density distribution the atomic gas associated with
Masks 2, 3, and 4. 
%
%
We then used Equation~(\ref{eq:3}) to estimate the contribution from
CNM gas to the observed [C\,{\sc ii}] emission in these Mask regions.
Note that
for a given thermal pressure, the intensity of [C\,{\sc ii}]
associated with H\,{\sc i} has a only weak dependence on the assumed
kinetic temperature, varying by about $\pm$20\% in the range between 60 and
200\,K.

 In Figure~\ref{fig:contributions} we show
the resulting contribution from atomic gas to the [C\,{\sc ii}]
intensity as a function of the Galactocentric distance.  We estimate
that the contribution from [C\,{\sc ii}] emerging from the atomic gas
component represents 21\% of the observed intensity.

\subsection{Ionized gas}
\label{sec:ionized-gas}

It has been suggested that a significant fraction of the [C\,{\sc ii}]
emission in the Galaxy observed by COBE emerges from the Extended Low
Density Warm Ionized Medium \citep[ELDWIM;][]{Petuchowski1993,Heiles1994}.  We
estimated the contribution to the observed [C\,{\sc ii}] emission from
an ionized gas using the Galactic distribution of the electron density
used by the NE2001 code \citep{Cordes2002}.  The NE2001 code assumes a
Galactic distribution of the electron density, including a thin and
thick disk, spiral arms (four), and Galactic center components. This
electron density distribution is further constrained using
observations of the dispersion measure of pulsars.

We used the NE2001 model to extract the electron density as a function
of heliocentric distance for all GOT\,C+ lines--of--sight in the
Galactic plane. For each step in heliocentric distance we also
calculate the column density of electrons and, assuming the [C]/[H]
gradient (Equation~\ref{eq:2}), the column density of
C$^+$ associated with the ionized gas.  We estimate the [C\,{\sc ii}]
intensity as a function of heliocentric distance for each
line--of--sight using Equation~(\ref{eq:3}), assuming a kinetic
temperature of 10$^4$\,K, and excitation by collisions with electrons
($R_{ul}=4.8\times10^{-8}$\,s$^{-1}$\,cm$^{-3}$;
\citealt{Wilson2002}). The term involving the kinetic temperature in
Equation~(\ref{eq:3}) becomes unity for such a large temperature,
therefore the intensity of [C\,{\sc ii}] associated with ionized gas
is insensitive to the kinetic temperature.  Note that the NE2001 model
is smooth and can only reproduce the electron density distribution
over large scales.  Finally, we calculated the distribution of the
[C\,{\sc ii}] emissivity as a function of Galactocentric distance
taking the same considerations as described in Appendix A. The
resulting radial distribution is shown in
Figure~\ref{fig:contributions}. We estimated that the contribution
from [C\,{\sc ii}] arising from ionized gas represents only 4\% of the
total observed intensity.

 \citet{Velusamy2012} observed an excess of [C\,{\sc ii}] emission at
the tangent velocities towards the Scrutum--Crux spiral arms tangent
which was interpreted as coming from compressed ionized gas.  They
suggested that this compression of ionized gas is the result of the
passing of a spiral arm density wave. Note, however, that this excess
can only be detected toward spiral arm tangencies due to a
projection effect by which the path length toward a spiral arm is
maximized. We therefore do not expect a significant contribution from
this excess to the overall observed [C\,{\sc ii}] intensity.

Note that our estimation of the contribution from ionized gas to
the observed [C\,{\sc ii}] emission corresponds only to that for
lines--of--sight in the Galactic plane with b=0\degr. The ELDWIM gas
is known to have a scale height of about 1\,kpc
\citep{Kulkarni1987,Reynolds1989}, which is much larger than that of
the atomic and molecular gas ($\sim100-300$\,pc). Thus, the
contribution from ionized gas to the total [C\,{\sc ii}]  {\it luminosity} of
the Galaxy should be larger than what is suggested by the emissivities
at $b=0$\degr. At a distance from the Sun of 4\,kpc, a 1\,kpc scale
height corresponds to 14\degr. Thus, within the a 7\degr\ beam of COBE,
the [C\,{\sc ii}] emission from the ionized component fills the beam,
while the emission from the Galactic plane (vertical angular scale
$\sim$2\,deg; \citealt{Nakagawa1998}) will be significantly beam
diluted.  The [C\,{\sc ii}] emission from the Galactic plane and
that from the extended ionized medium should appear comparable within
a COBE beam, as suggested by several analyses
\citep{Petuchowski1993,Heiles1994}. A detailed analysis of the relationship
between the {\it Herschel}/HIFI [C\,{\sc ii}] observations and those from
COBE/FIRAS will be presented in a future paper.  

\subsection{Dense photon-dominated regions}
\label{sec:dense-phot-domin}

\begin{figure}[t]

  \centering
   \includegraphics[width=0.5\textwidth]{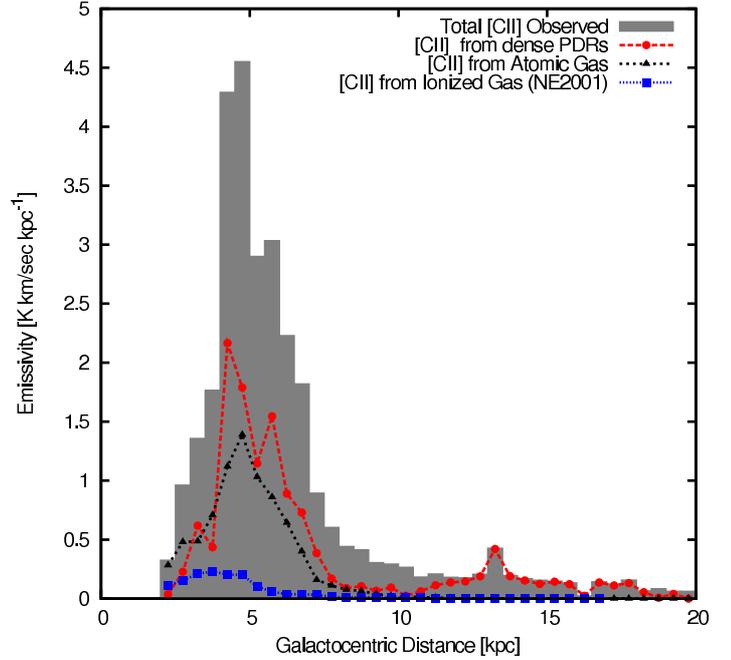}
   \caption{Azimuthally--averaged radial distribution of the observed
     [C\,{\sc ii}] emissivity in Masks\, 2, 3, and 4 together with the
     different contributions from the ISM components discussed in
     Section~\ref{sec:origin-c-sc}. Typical uncertainties are
     0.02\,K\,km\,s$^{-1}$\,kpc$^{-1}$ for the total [C\,{\sc ii}] emission
     and that contributed from dense PDRs, and
     0.001\,K\,km\,s$^{-1}$\,kpc$^{-1}$ for the contribution from H\,{\sc
     i} gas. For the contribution from ionized gas, we assume 30\%
     uncertainties or 0.07\,K\,km\,s$^{-1}$\,kpc$^{-1}$.
}
\label{fig:contributions}
\end{figure}

It is possible that a significant fraction of the observed [C\,{\sc
ii}] intensity in the Galactic plane comes from dense PDRs, as
observations of Galactic and extragalactic PDRs have shown that they
are the sources of bright [C\,{\sc ii}] emission
\citep[e.g.][]{Stacey1991, Boreiko1988}.  Note that the main--beam
temperatures in the GOT\,C+ survey are mostly in the 0.5--8\,K range,
with only a few spaxels with $T_{\rm mb}>$8\,K (see Fig. 7 in
\citealt{Goldsmith2012}). This range of main beam temperatures is well
below that observed in nearby star--forming regions such as Orion
($\sim$100\,K) which are powered by the intense FUV radiation field
produced by nearby massive stars.  Bright PDRs are, however, not
numerous enough to dominate the global [C\,{\sc ii}] emission in the
Milky Way.  In Section~\ref{sec:fuv-radiation-field} (see also
\citealt{Pineda2010a}) we found that most of the [C\,{\sc ii}]
emission in the Galactic plane arises from regions which are exposed
to moderate--strength FUV fields, in agreement with models of the
Galaxy based on COBE data \citep{Cubick2008}.  We considered [C\,{\sc
ii}] velocity components associated with $^{13}$CO (Mask\,4) as likely
to be associated with dense and warm PDRs.  Such regions have large
enough column densities, and presumably sufficient volume densities,
to produce detectable $^{13}$CO emission while they are also warm
enough to produce detectable [C\,{\sc ii}] emission.  We determined
the contribution from dense PDRs to the observed [C\,{\sc ii}]
intensity in Mask\,4 by subtracting the [C\,{\sc ii}] emission
associated with atomic gas, as determined in
Section~\ref{sec:atomic-gas}.  The resulting radial distribution of
the [C\,{\sc ii}] emissivity associated with dense PDRs is shown as a
red-dashed line in Figure~\ref{fig:contributions}.  The [C\,{\sc
ii}] emission emerging from dense PDRs represents 47\% of the total
emission observed in the GOT\,C+ survey.

\subsection{[C\,{\sc ii}] Excess}
\label{sec:cii_excess}

The combined contribution to the [C\,{\sc ii}] emission from atomic
gas, ionized gas, and dense PDRs, as estimated above, can account for
only 72\% of the observed emission.  It has been suggested the
remaining [C\,{\sc ii}] emission is produced in the envelopes of dense
molecular clouds where the H\,{\sc i}/H$_2$ transition is largely
complete, but the column densities are low enough that the
C$^{+}$/C$^{0}$/CO transition is not
\citep[e.g][]{Madden1997,Langer2010}. Since this H$_2$ gas component
is not traced by CO, but by [C\,{\sc ii}] and to a lesser extent by
[C\,{\sc i}], we refer to it as ``CO--dark H$_2$ gas''.  In the
following section we derive the distribution of the CO--dark
H$_2$ gas in the plane of the Milky Way, and compare it to that traced
by $^{12}$CO and $^{13}$CO.

%

\begin{figure}[t]
  \centering
   \includegraphics[width=0.45\textwidth]{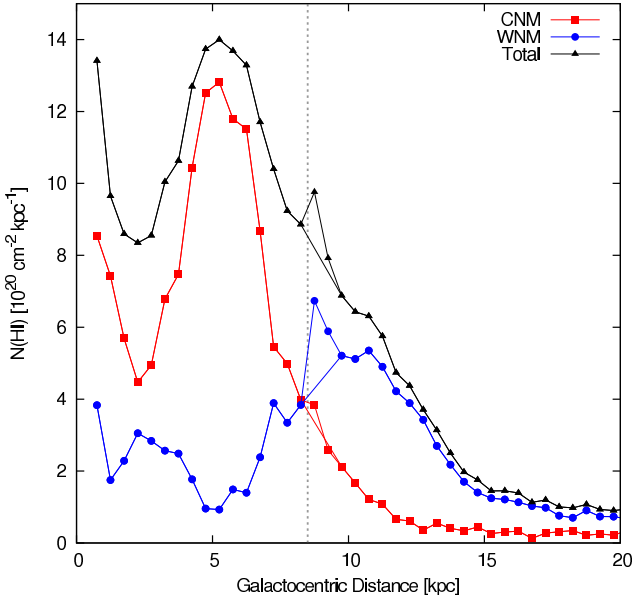}
      \caption{Radial distribution of the azimuthally averaged total
      H\,{\sc i} column density in the Galactic plane. We also include
      the estimated radial distribution of the cold neutral medium
      (CNM) and warm neutral medium (WNM) gas.  Typical uncertainties
      of the H\,{\sc i} column densities are
      10$^{18}$\,cm$^{-2}$\,kpc$^{-1}$. The vertical line shows the
      Solar radius. At 9\,kpc there is a discontinuity in the total
      and WNM $N$(H\,{\sc i}) distribution which is attributed to the
      abrupt change between the number of rings sampled in the inner
      and outer Galaxy. We therefore also show lines that smoothly
      connect the distributions in the inner and outer Galaxy. }
\label{fig:cnm_wnm_colden}
\end{figure}

\begin{figure}[t]
  \centering
   \includegraphics[width=0.45\textwidth]{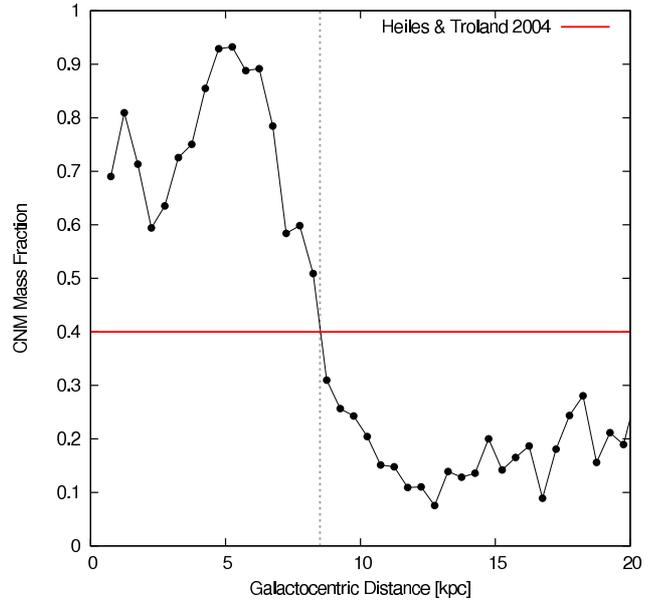}
      \caption{Radial distribution of the mass fraction of the atomic
gas in the cold neutral medium (CNM). The horizontal line shows the
local mass fraction of the CNM derived by \citet{Heiles2003}. The
vertical line denotes the solar radius. The typical uncertainty in the
ratio is $\sim$0.002.}
\label{fig:wnm_cnm_fraction}
\end{figure}

\section{The distribution of the ISM components in the plane of the Milky Way}
\subsection{The Warm and Cold Neutral Medium}
\label{sec:disent-cnm-wnm}

As mentioned earlier, atomic gas is predicted to be in two phases in
roughly thermal pressure equilibrium, the Cold Neutral Medium (CNM;
$n_{\rm H}\simeq50$\,cm$^{-3}$; $T_{\rm kin}\simeq$80\,K) and the Warm
Neutral Medium (WNM; $n_{\rm H}\simeq 0.5$\,cm$^{-3}$; $T_{\rm
kin}\simeq$8000\,K).  Because atomic hydrogen is one of the main
collisional partners of C$^+$, the [C\,{\sc ii}] line is a tracer of
atomic gas that has the advantage over the 21\,cm line (observed in
emission) that it is sensitive to both the volume density and the
kinetic temperature of the gas (Equation~\ref{eq:3}). Thus, the
[C\,{\sc ii}] line should in principle trace both the WNM and CNM
atomic gas components.  In practice, however, because of the volume
density contrast between the CNM and WNM, the [C\,{\sc ii}] line is
expected to be significantly weaker for WNM gas compared to the
emission from CNM gas \citep{Wolfire2010}.  For the H\,{\sc i} peak
intensity of 160K (in a 0.8 km s$^{-1}$ channel width) at the 5\,kpc
ring, and considering the abundance and thermal pressure gradients
(Equation~\ref{eq:2} and \ref{eq:4}, respectively), we can use
Equation~(\ref{eq:3}), assuming 100\,K for the CNM and 8000\,K for the
WNM, to estimate a [C\,{\sc ii}] intensity associated with H\,{\sc i}
of 0.3K in the case that all gas is CNM, and of 0.04K in case all gas
is WNM. Considering our 3$\sigma$ sensitivity limit per channel of
0.3\,K we find that weak CNM gas can be detected, while WNM gas would
be too weak to be detected in individual spaxels. 

 When azimuthally averaging (Section~\ref{sec:atomic-gas}), however,
[C\,{\sc ii}] emission associated with CNM can produce a significant
contribution to the observed [C\,{\sc ii}] emissivity. For the $R_{\rm
gal}=5$\,kpc peak of $1.4\times10^{21}$\,cm$^{-2}$\,kpc$^{-1}$, we
find a contribution of 1.1\,K\,km\,s$^{-1}$\,kpc$^{-1}$ for all
H\,{\sc i} gas in CNM, and 0.04\,K\,km\,s$^{-1}$\,kpc$^{-1}$ for all
H\,{\sc i} gas in WNM. As the 3$\sigma$ sensitivity limit for the
azimuthal average is 0.06\,K\,km\,s$^{-1}$\,kpc$^{-1}$
(Appendix~\ref{sec:determ-azim-aver}), we see that WNM is always below
the sensitivity limit even for the azimuthal average. This suggests
that only the CNM can contribute to the observed [C\,{\sc ii}]
emission.  We therefore assume that spaxels that have both H\,{\sc i}
and [C\,{\sc ii}] emission (Masks 2, 3, and 4) are associated with CNM
gas. In our survey, there are also spaxels where H\,{\sc i} and
$^{12}$CO emission are detected but [C\,{\sc ii}] emission is not
(Mask\,1). These regions are likely too cold and/or have too low
volume densities for [C\,{\sc ii}] to be detected, but can still be
associated with cold H\,{\sc i} gas. We therefore also consider
H\,{\sc i} associated with Mask\,1 as being in the CNM phase. Finally,
we consider spaxels with H\,{\sc i} but neither [C\,{\sc ii}] nor
$^{12}$CO emission (Mask\,0) to be associated with WNM gas. In the
following we use this criteria to separate the CNM and WNM components
from the observed H\,{\sc i} emission and study their distribution in
the plane of the Milky Way.

It is possible that some spaxels have [C\,{\sc ii}] emission that is
below the sensitivity limit for individual spaxels but can still be
detected in the azimuthal average. In
Figure~\ref{fig:mask_radial_distro}, we see that [C\,{\sc ii}]
emission ($\sim$0.1\,K\,km\,s$^{-1}$\,kpc$^{-1}$) in the inner galaxy
can be detected in the region where only H\,{\sc i} is detected in the
pointed observations (Mask\,0).  The average H\,{\sc i} column for
$R_{\rm gal}=3-8$\,kpc ($\sim1\times10^{20}$\,cm$^{-2}$\,kpc$^{-1}$),
would imply a [C\,{\sc ii}] emissivity of about
0.17\,K\,km\,s$^{-1}$\,kpc$^{-1}$ in the case that all gas is CNM and
0.006\,K\,km\,s$^{-1}$\,kpc$^{-1}$ in the case that all gas is
WNM. Thus, the emission detected in Mask\,0 can only be associated
with CNM gas.  Note that for the inner galaxy, when we associate
H\,{\sc i} with [C\,{\sc ii}] and/or $^{12}$CO emission in individual
spaxels, we are assuming that the H\,{\sc i} and [C\,{\sc ii}] and/or
$^{12}$CO emission arises from both the near and far side of the
Galaxy.  We could in principle overestimate the H\,{\sc i} column
density associated with the CNM in the case that the [C\,{\sc ii}]
and/or $^{12}$CO emission arises from either the near or far side of
the galaxy, but H\,{\sc i} from both sides. We consider this situation
unlikely, however, as we expect that the physical conditions within a
galactocentric ring to be relatively uniform. The maximum effect, in
the case that all LOSs in the inner galaxy are affected, is a
reduction of the azimuthally averaged CNM column density by a factor
of $\sim$2 (and a corresponding increase of the WNM column density by
the same factor).

In Figure~\ref{fig:cnm_wnm_colden}, we show the radial distribution of
the total H\,{\sc i} column density together with that of the
contributions from CNM and WNM gas.  The column densities are
estimated from the H\,{\sc i} emissivity distributions which are in
turn calculated considering all spaxels with H\,{\sc i} emission as
well as those associated with CNM gas (Masks 1, 2, 3, and 4) and those
associated with WNM gas (Mask\,0).   The column density of CNM gas
was corrected for opacity effects as described in
Section~\ref{sec:atomic-gas}. The value of $N$(H\,{\sc i}) at $R_{\rm
gal}=8.5$\,kpc ($6.7\times10^{20}$\,cm$^{-2}$\,kpc$^{-1}$; without the
opacity correction) is in good agreement with the local value of the
H\,{\sc i} surface density of $6.2\times10^{20}$\,cm$^{-2}$ found by
\citet{Dickey1990}.  As we can see, the CNM dominates the H\,{\sc i}
column density in the inner Galaxy, while WNM dominates in the outer
Galaxy. The column density of the WNM gas in the inner Galaxy is
similar to what we estimated for the H\,{\sc i} column density
required to reproduce the [C\,{\sc ii}] emission resulting from the
azimuthal average in Mask 0. This result indicates that even a larger
fraction of the atomic gas in the inner Galaxy might be in the form of
CNM.  CNM clouds may still exist in the outer Galaxy, as molecular
clouds and star formation are present, and the reduction of its
contribution in the outer Galaxy might be a result of a reduced
filling factor of the CNM gas.

 There is a discontinuity in the total H\,{\sc i} column density
distribution at around 9\,kpc, which may be an artifact of the assumed
flat rotation curve with purely circular motions. An alternative
explanation is shown in Figure~\ref{fig:different_bins} in
Appendix~\ref{sec:determ-azim-aver}, which shows the number of ring
samples as a function of Galactocentric distance.  Because a LOS
samples a ring in the inner Galaxy twice but a ring in the outer
Galaxy only once, there is a discontinuity in the number of samples
between the inner and outer Galaxy.  This discontinuity will therefore
affect the azimuthally averaged emissivity distributions. The effect
on the [C\,{\sc ii}] and $^{12}$CO emissivity distributions is small
because it coincides with a steep decrease in the value of the sum of
the emissivities for each sample (see numerator in
Equation~\ref{eq:5}). In the case of H\,{\sc i}, however, the sum of
the emissivities for each sample is smooth at around 9\,kpc, and
therefore dividing by the number of samples results in a noticeable
discontinuity. In Figure~\ref{fig:cnm_wnm_colden}, we also show the
H\,{\sc i} column density distribution where data points with $R_{\rm
gal} < 9$\,kpc and $R_{\rm gal} > 11$\,kpc are smoothly connected.

In Figure~\ref{fig:wnm_cnm_fraction} we present the mass fraction of
the CNM gas. In the inner Galaxy, up to 90\% of the atomic gas is in
the form of CNM, with even a larger percentage being possible based on
the discussion above. For larger Galactocentric distances, this
fraction decreases to only 0.1--0.2, showing little variation with
Galactocentric distance.  These results are in good agreement with
those found by \citet{Dickey2009} using H\,{\sc i} emission/absorption
observations in the outer Galaxy.  At $R_{\rm gal}=8.5$\,kpc, we find
that the fraction is in good agreement with the local fraction of 0.4
derived by \citet{Heiles2003}. We derive an average CNM mass fraction
for the entire Galactic disk of 0.43.

Our results suggest that the CNM is the dominant atomic gas phase in
the inner Galaxy while the WNM dominates the outer Galaxy. The
increase of the fraction of CNM gas towards the inner Galaxy coincides
with that of the $^{12}$CO and [C\,{\sc ii}] emissivities. This
suggests that the CNM plays a important role as the precursor of
molecular clouds that will eventually form stars.

\subsection{Molecular Hydrogen}
\label{sec:distr-h_2-plane}

%

   \begin{figure}[t]
  \centering
   \includegraphics[width=0.35\textwidth]{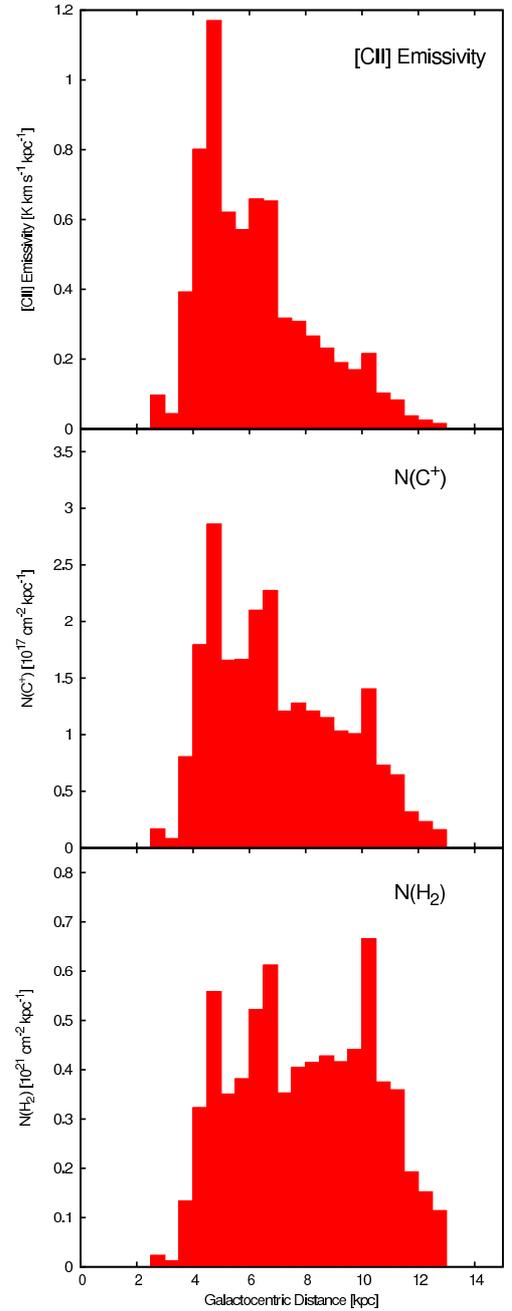}
      \caption{Radial distribution of the azimuthally averaged excess
      [C\,{\sc ii}] emissivity as discussed in
      Section~\ref{sec:cii_excess} ({\it upper panel}).  Radial
      distribution of the C$^+$ column density derived from the excess
      [C\,{\sc ii}] emissivity ({\it middle panel}). Radial
      distribution of the H$_2$ column densities derived from $N({\rm
      C}^+)$ associated with the excess [C\,{\sc ii}] emissivity ({\it
      lower pannel}).}
\label{fig:temperature_pressure}
\end{figure}

\subsubsection{CO--dark H$_2$ gas}
\label{sec:distr-dark-h_2}

To derive the column density radial distribution of the CO--dark H$_2$
gas component, we first estimated the C$^+$ column density associated
with the [C\,{\sc ii}] excess emission derived in
Section~\ref{sec:origin-c-sc}. The C$^+$ column density, which can be
derived by inverting Equation~(\ref{eq:3}), depends on the observed
[C\,{\sc ii}] intensity, the H$_2$ volume density, $n_{\rm H_2}$, and
the kinetic temperature of the H$_2$ gas, $T^{\rm H_2}_{\rm
kin}$. Without an independent estimate of the physical conditions in
the CO--dark H$_2$ layer, we need to make an assumption about the
value of $n_{\rm H_2}$ and $T^{\rm H_2}_{\rm kin}$. In
Section~\ref{sec:atomic-gas}, we derived the contribution from atomic
gas (CNM) to the observed [C\,{\sc ii}] intensity by estimating the
H\,{\sc i} volume density, $n_{\rm HI}$, from the radial thermal
pressure distribution (Equation~\ref{eq:4}), assuming a kinetic
temperature of the H\,{\sc i} gas, $T^{\rm HI}_{\rm kin}$, of 100\,K.
After the H/H$_2$ transition takes place, and as the column density
increases, self--gravity starts to become important, resulting in a
gradual increase of the volume density of the gas. The increased
shielding of FUV photons results in a decreased gas heating, while
line cooling is largely unaffected due to the larger volume densities,
resulting in a reduction of the kinetic temperature.   Overall,
however, the thermal pressure of the gas increases with column
density: studies of local clouds, in CO absorption in the millimeter
\citep{Liszt1998} and in the UV (Goldsmith 2013 submitted), tracing
column densities up to that of the C$^+$/C$^0$/CO transition layer,
typically suggest thermal pressures in the range between 10$^2$ to
10$^{4}$ K\,cm$^{-3}$, while for well shielded regions, traced by
$^{13}$CO, excitation studies suggest thermal pressures in the range
between 10$^4$ to 10$^5$ K\,cm$^{-3}$ \citep{Sanders1993}.   It is
therefore likely that a thermal pressure gradient is present in the
region between the H/H$_2$ and C$^+$/C$^0$/CO transition layers, where
the emission from CO--dark H$_2$ gas originates. To estimate the
column density of CO-dark H$_2$ gas, we assume that the thermal
pressure of the CO--dark H$_2$ gas is the geometric mean between that
of the H\,{\sc i} gas (Equation~\ref{eq:4}) and the maximum observed
at the C$^+$/C$^0$/CO transition layer. At the solar neighborhood, the
minimum pressure given by Equation~\ref{eq:4} is 3000 K\,cm$^{-3}$
while the maximum pressure at the C$^+$/C$^0$/CO transition layer
given by Goldsmith (2013) is 7400 K\,cm$^{-3}$.  The resulting
geometric mean is 4700 K\,cm$^{-3}$, which is a factor of 1.6 larger
than the thermal pressure given by Equation~(\ref{eq:4}). We therefore
multiply Equation~(\ref{eq:4}) by this factor to determine the mean
thermal pressure of the CO--dark H$_2$ gas as a function of
Galactocentric distance.  We assume a constant kinetic temperature of
$T^{\rm H_2}_{\rm kin}$=70\,K, following the results from
\citet{Rachford2002}, and a corresponding
$R_{ul}=4.3\times10^{-10}$\,s$^{-1}$\,cm$^{-3}$ \citep{Flower1977} for
collisions with H$_2$. Using this kinetic temperature we derive the
H$_2$ volume density of the CO--dark H$_2$ gas from
Equation~(\ref{eq:4}) as discussed above.  We will discuss below the
effects of choosing different values of $T^{\rm H_2}_{\rm kin}$ and
$n({\rm H}_2)$ in the derivation of the column density of the CO--dark
H$_2$ gas.  We finally converted the radial distribution of the C$^+$
column density to that of the H$_2$ column density, assuming that all
gas--phase carbon is in the form of C$^+$, by applying a [C]/[H$_2$]
abundance gradient. The [C]/[H$_2$] abundance ratio is twice the
[C]/[H] ratio given in Equation~(\ref{eq:2}).

   \begin{figure}[t]
  \centering
   \includegraphics[width=0.48\textwidth]{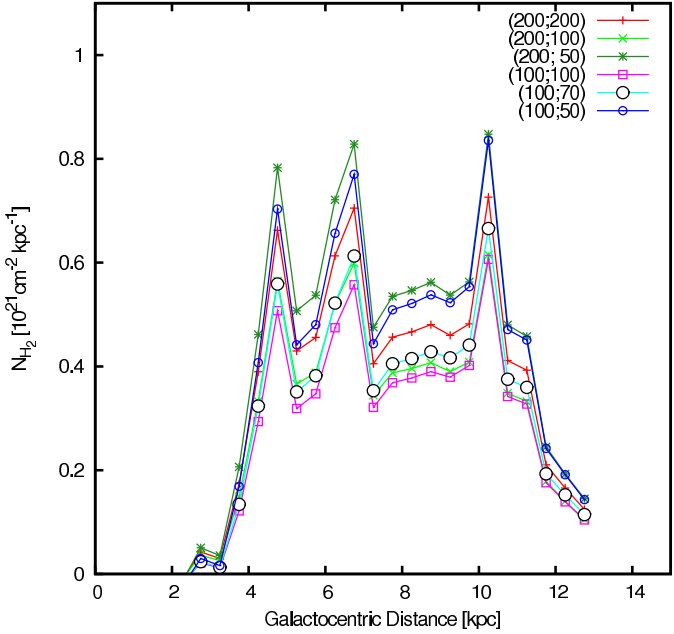}
      \caption{CO--dark H$_2$ gas column density as a function of
              Galactocentric distance for different H\,{\sc i} and
              H$_2$ temperature pairs ($T^{\rm HI}_{\rm kin}$;$T^{\rm
              H_2}_{\rm kin}$). The corresponding volume densities
              were derived from Equation~(\ref{eq:4}; see Section~\ref{sec:distr-dark-h_2}).}
\label{fig:different_temps}
\end{figure}

In Figure~\ref{fig:temperature_pressure}, we show the
azimuthally--averaged radial distribution of the [C\,{\sc ii}] excess
emissivity (top panel), of the corresponding C$^+$ column density
(middle panel), and of the CO--dark H$_2$ column density (lower
panel).  The steep distribution of the [C\,{\sc ii}] excess
emissivity, with a peak at 4.2\,kpc, becomes flatter when converted to
$N({\rm C}^+)$ as a result of applying the $n({\rm H}_2)$ gradient
derived from Equation~(\ref{eq:4}) and a constant $T^{\rm H_2}_{\rm
kin}$ equal to $70$\,K. When applying the [C]/[H$_2$] abundance
gradient, the radial distribution of CO--dark H$_2$ becomes nearly
constant between 4 and 11\,kpc.  We will discuss this flat
distribution of the CO--dark H$_2$ column density and relate it to the
denser H$_2$ gas traced by CO and $^{13}$CO in
Section~\ref{sec:compl-distr-h_2}.   Note that when calculating
column densities we are assuming that the [C\,{\sc ii}] emission is
extended with respect to the beam of our observations. In the case
that there is substructure within the beam, our estimate of the column
density will be a lower limit, as it corresponds to that averaged over
{\it Herschel}'s 12\arcsec\ beam.

In Figure~\ref{fig:different_temps} we show the resulting CO--dark
H$_2$ column density distribution for different combinations of the
kinetic temperature in the H\,{\sc i} and H$_2$ layers.  The volume
densities are obtained from Equation~(\ref{eq:4}), considering the
factor 1.6 correction for H$_2$ gas discussed above.  We assume that
the temperature of the H$_2$ layer is equal to or lower than the
temperature of the H\,{\sc i} layer.  For each kinetic
temperature, the corresponding value of $R_{ul}$ was interpolated from
data calculated by \citet{Flower1977} for collisions with H$_2$ and
by \citet{Launay1977} for collisions with H.  As we can see, for a wide range of
temperature combinations, the derived column density deviates by about
30\% from our adopted model with $T_{\rm kin}^{\rm HI}=100$\,K and
$T^{\rm H_2}_{\rm kin}=70$\,K. We therefore consider that our CO--dark
H$_2$ column density determination has a 30\% uncertainty.

\subsubsection{ H$_2$ gas traced by $^{12}$CO and $^{13}$CO}
\label{sec:h_2-gas-traced}

A commonly employed method to estimate the H$_2$ mass of molecular
clouds is the use of the $^{13}$CO integrated intensity, $I(^{13}{\rm
CO})$, to derive the $^{13}$CO column density, $N(^{13}{\rm CO})$,
assuming that this line is optically thin and that local thermodynamic
equilibrium (LTE) applies.  $N(^{13}{\rm CO})$ can be then converted
to $N({\rm CO})$, assuming an appropriate isotopic ratio, and $N({\rm
CO})$ to $N({\rm H}_2)$, assuming a CO abundance relative to H$_2$.
It has been observed in large--scale $^{12}$CO and $^{13}$CO surveys,
however, that the spatial extent of the $^{13}$CO emission is often
noticeably smaller than, and is contained within, the observed extent
of the $^{12}$CO emission \citep{Goldsmith2008,Roman-Duval2010}. This
spatial discrepancy is mostly due to the limited sensitivity of the
observations that is often inadequate to detect the weak $^{13}$CO
emission emerging from the envelopes of molecular clouds.  The region
where only $^{12}$CO is detected but $^{13}$CO is not can account for
$\sim$30\% of the total mass of a molecular cloud
\citep{Goldsmith2008,Pineda2010a}. The discrepancy between the spatial
extent of these two lines is also seen in the position velocity maps
of the observed $^{12}$CO and $^{13}$CO lines obtained as part of the
GOT\,C+ survey.

\begin{figure}[t]
  \centering
   \includegraphics[width=0.45\textwidth]{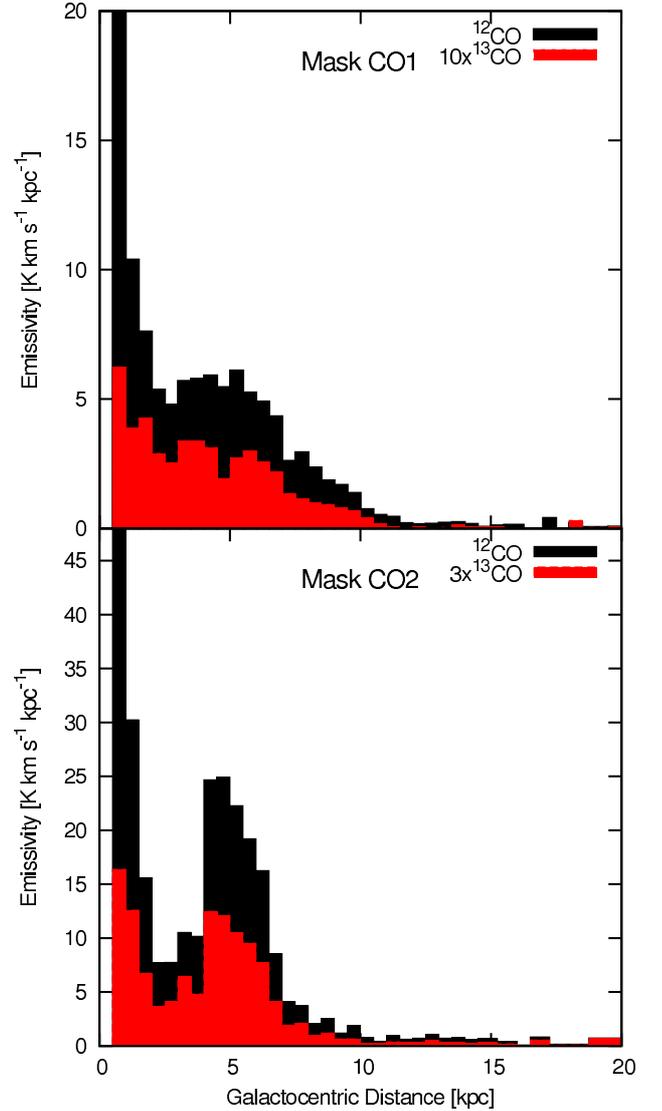}
      \caption{Radial distribution of the azimuthally averaged
$^{12}$CO and $^{13}$CO emissivities observed in the Galactic plane for
the case where $^{12}$CO is detected in individual spaxels but
$^{13}$CO is not (Mask CO1; upper panel), and for the case where both
$^{12}$CO and $^{13}$CO are detected in individual spaxels (Mask CO2;
lower panel). Typical uncertainties of the $^{12}$CO and $^{13}$CO
emissivities are  0.1 and 0.02\,K\,km\,s$^{-1}$\,kpc$^{-1}$, respectively. }
\label{fig:masks_distro}
\end{figure}

 To estimate the azimuthally averaged H$_2$ column density traced by
 CO and $^{13}$CO for each ring, we considered two cases: spaxels
 where both $^{12}$CO and $^{13}$CO are detected (Mask CO2) and
 spaxels where $^{12}$CO is detected but $^{13}$CO is not (Mask
 CO1). In Figure~\ref{fig:masks_distro}, we show the radial
 distribution of the azimuthally averaged $^{12}$CO and $^{13}$CO
 emissivity for spaxels in Masks CO1 and CO2.  Although by definition
 spaxels in Mask\,CO1 have $^{13}$CO intensities below the sensitivity
 limit, we could nevertheless detect $^{13}$CO emission for spaxels in
 this mask region after averaging them in azimuth as seen in the
 figure.  As $^{13}$CO can be considered optically thin, the
 integrated intensity of this line is proportional to the $^{13}$CO
 column density which in turn should be proportional to the H$_2$
 column density.  Thus, spaxels in Masks CO2 and CO1 correspond to
 regions with larger and smaller column densities, respectively, with
 the threshold between these two column density regimes given by the
 sensitivity of the $^{13}$CO observations.

 In the following we describe our procedure used to estimate
 H$_2$ column densities from Mask\,CO1 and Mask\,CO2. The procedure is
 very similar to that used in the Taurus molecular cloud by
 \citet{Pineda2010a} and \citet{Goldsmith2008}. We refer to these
 papers for further details on the column density determination.  We
 derived the H$_2$ column density using the $^{12}$CO and $^{13}$CO
 emissivities calculated from spaxels where the $^{12}$CO and/or
 $^{13}$CO are detected.  We later expressed the derived H$_2$ column
 density as an average that considers  all sampled spaxels within a given
 Galactocentric ring, not only where $^{12}$CO and $^{13}$CO emission
 is detected, to facilitate the comparison with the column density of
 CO--dark H$_2$ gas (Section~\ref{sec:distr-dark-h_2}).

 For Mask\,CO2, we estimated $N(^{13}{\rm CO})$ from $I(^{13}{\rm
   CO})$ under the assumption that this line is (almost) optically
   thin and that LTE applies.  Apart from the dependence on
   $I(^{13}{\rm CO})$, $N(^{13}{\rm CO})$ depends on the excitation
   temperature, $T_{\rm ex}$, and opacity corrections that might
   apply. The excitation temperature and the optical depth can be
   derived from the peak line intensity of the $^{12}$CO and $^{13}$CO
   lines, respectively.  To derive the line peak intensities we
   assumed that the average $^{12}$CO or $^{13}$CO emissivities are
   equivalent to the average of the integrated intensity of a number
   of Gaussian lines with equal FWHM line width of 3\,km s$^{-1}$,
   which is a typical value for the observed $^{12}$CO and $^{13}$CO
   lines in our survey (Langer et al. 2013 in preparation).  Note that the calculation of $T_{\rm ex}$
   also assumes that the $^{12}$CO line is optically thick ($\tau>>1$). Accounting
   for opacity in the derivation of $N(^{13}{\rm CO})$ typically
   results in only a 10\% correction.

We converted $N(^{13}{\rm CO})$ to $N({\rm CO})$ using the slope of
the $^{12}$C/$^{13}$C isotope ratio gradient derived by
\citet{Savage2002}, while fixing the $^{12}$C/$^{13}$C isotope ratio
to be 65 at $R_{\rm gal}=8.5$\,kpc. The $^{12}$C/$^{13}$C isotope ratio
as a function of Galactocentric distance is thus given by

\begin{equation}
\label{eq:iso}
\frac{^{12}\rm C}{^{13} \rm C} =4.7 \left ( \frac{R_{\rm gal}}{{\rm kpc}} \right )+25.05,
\end{equation}
where $R_{\rm gal}$ is the Galactocentric distance.  For this
gradient, the isotopic ratio varies from $\sim$43 to $\sim$80 for
Galactocentric distance between 4\,kpc and 12\,kpc.  We converted
$N({\rm CO})$ to $N({\rm H}_2)$ using a [CO]/[H$_2$] abundance
gradient constructed using the slope of the [C]/[H] abundance gradient
in Equation~(\ref{eq:2}) fixing [CO]/[H$_2$]=$1\times10^{-4}$ at
$R_{\rm gal}=$8.5\,kpc. 

As noted by e.g. \citet{Tielens2005}, the [CO]/[H$_2$] relative
abundance used here, and widely used in the literature, is a factor of
3 lower than the gas-phase carbon abundance observed in the diffuse
ISM, used in Section~\ref{sec:distr-dark-h_2}. This reduced CO
fractional abundance is inconsistent with chemical models that predict
that, in well shielded regions, all the gas--phase carbon should be in
the form of CO \citep{Solomon1972,Herbst1973}.  This discrepancy is
not a result of CO molecules frozen onto dust grains, as it persist
even after a correction to account for CO and CO$_2$ ices is applied
\citep{Whittet2010,Pineda2010a}.  The CO/H$_2$ abundance is often
derived by comparing observations of CO and $^{13}$CO with visual
extinction maps towards well studied nearby dark clouds
\citep[e.g.][]{Dickman78,FLW1982,Pineda2010a}.  When the contribution
from H\,{\sc i} gas is subtracted, the visual extinction traces the
total column of H$_2$ along the line of sight. However, a certain
column of H$_2$ is required to shield the FUV photons before CO can
form. Therefore, for a given path length, only a fraction of it, its
central part, will have all gas--phase carbon in the form of CO, while
the remainder will be what we call the CO--dark H$_2$ gas.  Dividing
the column density of CO with that of H$_2$ will result in a
fractional abundance that is lower than what would result if CO is the
dominant form of carbon over the entire path length. This effect is
more prominent in regions with moderate column densities, which
represent the bulk of the mass in giant molecular clouds, where the
CO--dark H$_2$ and CO--traced layers are comparable.  Following this
argument, the local values, [C]/[H$_2$]=$2.8\times10^{-4}$ and
[CO]/[H$_2$]=$1\times10^{-4}$, would imply a CO--dark H$_2$ gas
fraction of 0.6, which is consistent with what is shown in
Figure~\ref{fig:nh2_fraction} and that independently found by
\citet{Paradis2012} studying the CO--dark H$_2$ fraction in the Solar
neighborhood using extinction mapping. Note that by using
[CO]/[H$_2$]=$1\times10^{-4}$ in our calculations we are implicitly
accounting for the contribution from CO--dark H$_2$ gas in the
derivation of the H$_2$ column density from CO and $^{13}$CO.  The
CO--dark H$_2$ gas present in regions traced by CO and $^{13}$CO will
be studied in detail by Langer et al. (2013 in preparation).

  As noted above, spaxels in Mask CO1 correspond to low H$_2$ column
  density gas. For molecular clouds it is likely that this gas is also
  characterized by low H$_2$ volume densities. Assuming that the
  detected $^{12}$CO emission in Mask CO1 is nonetheless optically
  thick we estimated the excitation temperature as we did for Mask
  CO2. Typical values were found to be in the 4\,K to 6\,K range for
  $4$\,kpc\,$\leq R_{\rm gal} \leq 12$\,kpc. As seen in Taurus, pixels
  in Mask CO1 are located in the periphery of the large column density
  molecular cloud traced by $^{12}$CO and $^{13}$CO
  (Mask\,CO2). However, the excitation temperatures estimated for Mask
  CO1 are lower than those derived for Mask CO2 ($\sim$10--20\,K), and
  therefore they are inconsistent with what we would expect for this
  gas being in LTE, because the kinetic temperature is likely to be
  higher in the outer regions of the cloud which are subject to
  increased heating.  It is therefore reasonable to assume that the
  gas in Mask CO1 is subthermally excited. \citet{Pineda2010a} used
  the RADEX \citep{vanderTak2007} excitation/radiative transfer code
  to predict line intensities from a gas with conditions similar to
  those of Mask\,CO1 in the envelope of the Taurus molecular cloud.
  RADEX accounts for effects of trapping, which are important for the
  excitation of CO at low densities.  The model parameters are the
  kinetic temperature, $T_{\rm kin}$, the H$_2$ volume density,
  $n({\rm H}_2)$, the CO column density per unit line width, $N({\rm
  CO})/\delta v$, and the $^{12}$CO/$^{13}$CO abundance ratio,
  $R$. The observed parameters are the excitation temperature and the
  $^{12}$CO/$^{13}$CO integrated intensity ratio.  As mentioned above,
  the excitation temperature varies from 6\,K to 4\,K for
  $4$\,kpc\,$\leq R_{\rm gal} \leq 12$\,kpc.  The $^{12}$CO/$^{13}$CO
  integrated intensity ratio shows little variation from its average
  value of $\sim$24 for the same range in Galactocentric distance.  We
  assume that the kinetic temperature is 15\,K. If the value of $R$
  were known, one could find a unique solution for $n({\rm H}_2)$ and
  $N({\rm CO})$. However, the isotopic ratio is expected to vary from
  that expected for well--shielded regions due to isotopic enhancements
  produced by  chemical and/or photo effects.  Thus, there is a family of
  solutions that can reproduce the observed
  quantities. \citet{Pineda2010a} used nearly 10$^6$ individual
  spectra in the Taurus molecular cloud to find a set of solutions that
  produced a monotonically decreasing $R$ with decreasing excitation
  temperature and a smoothly decreasing column density with decreasing
  excitation temperature.  The observed $^{12}$CO/$^{13}$CO integrated
  intensity ratio for a given value of $T_{\rm ex}$ for all
  Galactocentric rings is in excellent agreement with the same
  quantities observed in the envelope of the Taurus molecular
  cloud. We therefore used the solution presented in
  \citet{Pineda2010a} to derive the $N({\rm CO})$ column density in
  Mask CO1 for each Galactocentric ring. For our observed values of
  $T_{\rm ex}$ and the $^{12}$CO/$^{13}$CO integrated intensity ratio,
  their solution suggests a gas with $n({\rm
  H}_2)\simeq250$\,cm$^{-3}$ and $R\simeq30$.  We use their fit to the
  relationship between CO column density and the observed excitation
  temperature given by

 \begin{figure}[t]
  \centering
   \includegraphics[width=0.48\textwidth]{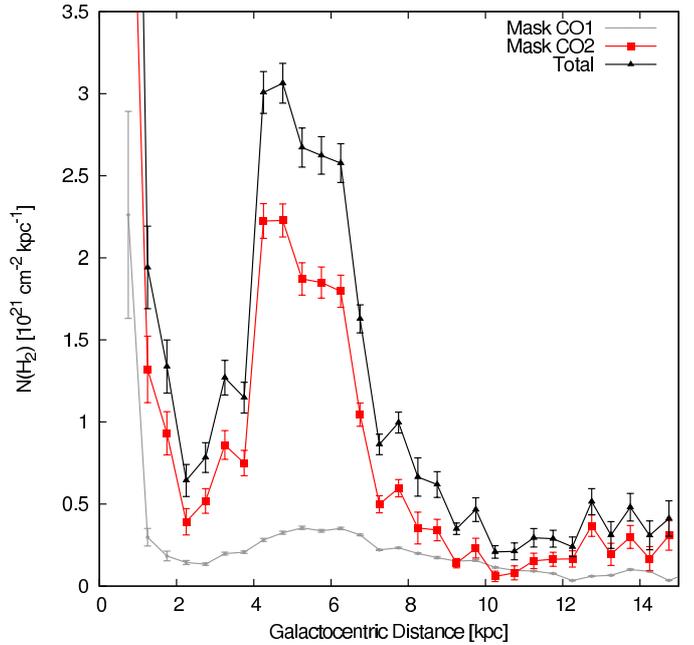}
      \caption{ Radial distribution of the azimuthally averaged H$_2$
column density as traced by $^{12}$CO and $^{13}$CO. We include the
radial distribution of $N({\rm H}_2)$ from gas associated with spaxels
where both $^{12}$CO and $^{13}$CO are detected (Mask CO2) and where
$^{12}$CO is detected but $^{13}$CO is not (Mask CO1). The total
distribution (Mask CO1+ Mask CO2) has been increased by 20\% to
account for $^{12}$CO emission that is not detected in individual
spaxels but was detected  after averaging them in azimuth (see
Section~\ref{sec:h_2-gas-traced}).  }
\label{fig:masks_nh2_distro}
\end{figure}

\begin{equation} 
     \left (  \frac{N({\rm CO})}{\rm cm^{-2}} \right ) \left ( \frac{\delta {\it v}}{\rm km\,s^{-1}}\right )^{-1} = 6.5\times10^{13} \left ( \frac{T_{\rm ex}}{{\rm K}} \right )^{2.7}
\end{equation} 
to derive the average $N({\rm CO})$ column density in Mask\,CO1.

 The [CO]/[H$_2$] abundance ratio is expected to show strong
differences between UV--exposed and shielded regions as a result of
the competition between the formation and destruction of CO
\citep[e.g.][]{vanDishBlack88,Visser2009}.  \citet{Pineda2010a}
compared CO column densities derived at the envelope of the Taurus
molecular cloud with a visual extinction, $A_{\rm V}$, map of the same
region (with the contribution from dust associated with atomic gas to
the visual extinction subtracted).  They provided a polynomial fit
giving $A_{\rm V}$ at any given value of $N({\rm CO})$ (see their
Figures 13 and 14).  We use this fit to derive visual extinctions from
$N({\rm CO})$ in Mask\,CO1.  Since $A_{\rm V}$ traces the total column
density of H$_2$, not only that traced by CO, the relationship between
$N({\rm CO})$ and $A_{\rm V}$ should already account for the CO--dark
H$_2$ gas present in the envelope of the Taurus molecular cloud. The
visual extinction outside the CO boundary in Taurus can be estimated
by searching for the value of $A_{\rm V}$ at which $N({\rm
CO})=0$\,cm$^{-2}$.  A polynomial fit to the lower part of the $N({\rm
CO})$--$A_{\rm V}$ relationship gives $N({\rm CO})=0$\,cm$^{-2}$ for
$A_{\rm V}$=0.37\,mag.   Note that this correction only applies if
the column density in Mask\,CO1 is derived using spaxels that have
both [C\,{\sc ii}] and $^{12}$CO emission. There are, however, spaxels
in Mask\,CO1 where $^{12}$CO emission is detected but [C\,{\sc ii}] is
not, and the correction in the conversion from $N({\rm CO})$ to
$A_{\rm V}$ does not apply to them, because their CO--dark H$_2$ layer
is not already traced by [C\,{\sc ii}].  We repeated the column
density estimation described above but for only spaxels in Mask\,CO1
that have both [C\,{\sc ii}] and $^{12}$CO emission as well as for
those that have $^{12}$CO emission but no [C\,{\sc ii}]. After
applying the 0.37\,mag correction to the column density distribution
derived from spaxels detected in [C\,{\sc ii}] and $^{12}$CO and
adding it to that derived from spaxels with $^{12}$CO emission only, we
find that the total column density distribution is about 15\% lower
than that without applying the correction. We therefore reduce the
derived values of $A_{\rm V}$ by 15\% to account for the CO--dark
H$_2$ gas already traced by [C\,{\sc ii}] in Mask\,CO1.

The formation of CO in cloud envelopes is mostly influenced by dust
grains shielding the FUV radiation field, with smaller contributions
from H$_2$ shielding and CO self--shielding. Thus, changes in
metallicity will not significantly influence the dust column density,
but will modify the H$_2$ column density, at which the C$^+$/C$^0$/CO
transition takes place, because a reduction/increase on metallicity is
associated with a increase/reduction of the gas--to--dust ratio,
$R_{\rm gd}= N({\rm H_2})$/$A_{\rm V}$.  To convert the derived visual
extinctions to H$_2$ column densities, we assumed that $R_{\rm gd}$
has the same slope, but with the opposite sign, of the galactic
metallicity gradient (Equation~\ref{eq:2}), and we fixed its value at
$R_{\rm gal}=8.5$\,kpc to be $9.4\times10^{20} {\rm cm}^{-2}\,{\rm
mag}^{-1}$ \citep{Bohlin1978}. The resulting gas to dust ratio as a
function of galactocentric distance is therefore given by,
\begin{equation}
R_{\rm gd}=2.38\times10^{20}10^{0.07R_{\rm gal}}\,\, {\rm cm}^{-2}/{\rm mag},
\end{equation}
with the galactocentric distance, $R_{\rm gal}$, in units of kpc.  We
applied this relationship to convert from $A_{\rm V}$ to $N({\rm
H_2})$ in Mask\,CO1.

\citet{Pineda2010a} and \citet{Goldsmith2008} detected $^{12}$CO and
$^{13}$CO emission in the periphery of Taurus where these lines were
not detected in individual pixels. \citet{Pineda2010a} estimated that
this gas accounts for about 20\% of the total mass of Taurus. Although
we can still detect weak $^{12}$CO emission after averaging in azimuth
regions where $^{12}$CO is not detected in individual spaxels, we could
not detect $^{13}$CO, and therefore cannot repeat the analysis done in
Taurus.  We corrected the total column density traced by $^{12}$CO and
$^{13}$CO by 20\% to account for this missing CO gas component.


Figure~\ref{fig:masks_nh2_distro} shows the total azimuthally averaged
H$_2$ column density as traced by $^{12}$CO and $^{13}$CO.  For the
range between 4\,kpc and 7\,kpc, gas associated with Mask\,CO2
contributes about 70\% of the total H$_2$ traced by CO and
$^{13}$CO. For 7\,kpc$ < R_{\rm gal} < $10\,kpc, the gas associated
with Mask CO2 contributes about 25\% of the total CO--traced H$_2$
gas. This contribution is consistent with the relative contribution
from Mask CO1 and Mask CO2 to the mass of the nearby Taurus molecular
cloud derived by \citet{Pineda2010a}.

\subsection{The complete distribution of H$_2$ in the plane of the Milky Way }
\label{sec:compl-distr-h_2}

\begin{figure}[t]
  \centering
   \includegraphics[width=0.45\textwidth]{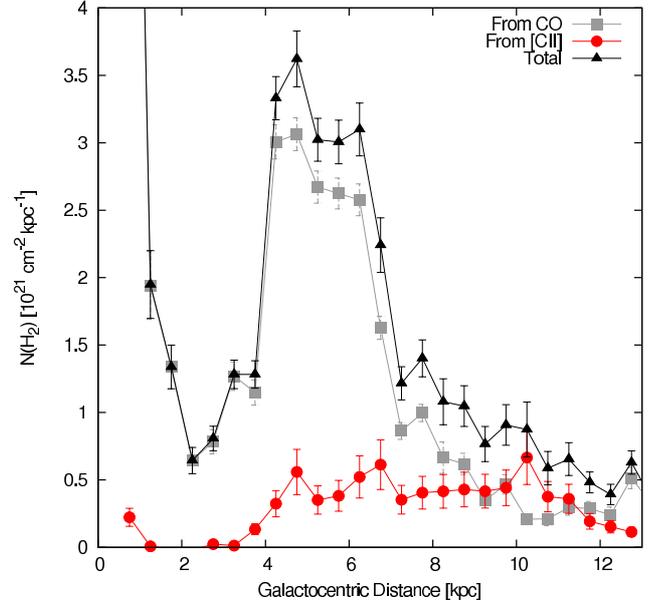}
      \caption{Radial distribution of the H$_2$ column  density
in the plane of the Milky Way. We also show the relative contributions
to the total H$_2$ column density from gas traced by
$^{12}$CO and $^{13}$CO and   from CO--dark
H$_2$ gas.}
\label{fig:co_cii_h2_column}
\end{figure}

\begin{figure}[t]
  \centering
   \includegraphics[width=0.45\textwidth]{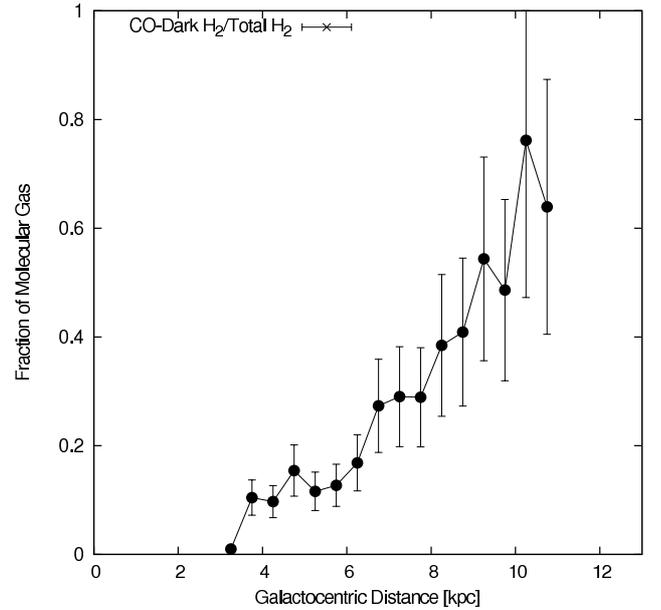}
      \caption{ Radial distribution of the fraction of the total H$_2$
      column density comprised by the CO--dark H$_2$ gas. }
\label{fig:nh2_fraction}
\end{figure}

In Figure~\ref{fig:co_cii_h2_column}, we show the radial distribution
of the total azimuthally averaged H$_2$ column density. The total
$N({\rm H}_2)$ is the combination of the contribution from H$_2$ gas
traced by CO and that traced by [C\,{\sc ii}] (CO--dark H$_2$
gas). Both contributions are also included in the figure.  Close to
the Galactic center, most of the H$_2$ is traced by CO with the
CO--dark H$_2$ gas making a negligible contribution. The CO--traced
H$_2$ gas is mostly concentrated in the 4\,kpc to 7\,kpc range, while
the CO--dark H$_2$ gas is extended over a wider range of
Galactocentric distances, between 4 and 11\,kpc.  As mentioned in
Section~\ref{sec:distr-dark-h_2}, the distribution of the CO--dark
H$_2$ column density is nearly constant between 4 and
11\,kpc. However, the fraction of the total molecular gas in this
component increases with Galactocentric distance. This increase is
shown in Figure~\ref{fig:nh2_fraction} where we show the CO--dark
H$_2$ gas fraction as a function of Galactocentric distance.  The
fraction of CO--dark H$_2$ gas increases monotonically with
Galactocentric distance, rising from 0.2 of the total H$_2$ gas at
4\,kpc to be about 0.8 at 10\,kpc.  The Galactic metallicity gradient
is likely accompanied by a decrease of the dust--to--gas ratio with
Galactocentric distance. The reduced dust extinction results in an
increased FUV penetration, which in turn results in the C$^+$/C$^0$/CO
transition layer taking place at larger H$_2$ column densities. Thus,
as suggested by our results, the low--column density H$_2$ gas at
large Galactocentric distances, with low--metallicities, is better
traced by [C\,{\sc ii}] (and perhaps [C\,{\sc i}]) than by CO.

For $R_{\rm gal}>11$\,kpc, we see a reduction of the column density of
CO--dark H$_2$ gas. In the outer Galaxy we detect fewer [C\,{\sc ii}]
clouds per unit area and, as we can see in
Figure~\ref{fig:contributions}, most of the detected emission is
associated with dense PDRs. Because of the reduced carbon abundance in the
outer Galaxy, and perhaps reduced thermal pressures, the [C\,{\sc ii}]
intensity associated with CO--dark H$_2$, as well as that associated
with H\,{\sc i} and electron gas, might be reduced to be below our
detection threshold, while only regions associated with
star--formation, dense PDRs, can produce detectable emission. Due to the
reduced star formation activity per unit area in the outer Galaxy, the
average strength of the FUV field is likely also reduced. This
reduction of the FUV field results in a lower kinetic temperature
that reduce the excitation of [C\,{\sc ii}] line.  Additionally, the
plane of the Milky Way is known to warp towards the outer Galaxy,
while the CO--dark H$_2$ distribution is based on [C\,{\sc ii}] data
at $b$=0\degr. Thus, our sampling of the outer Galactic plane might
not be as good as it is for the inner Galaxy. Therefore we cannot rule
out that the distribution of H$_2$ extends even further in the outer
Galaxy than indicated here.

   \begin{figure}[t] 
  \centering
   \includegraphics[width=0.45\textwidth]{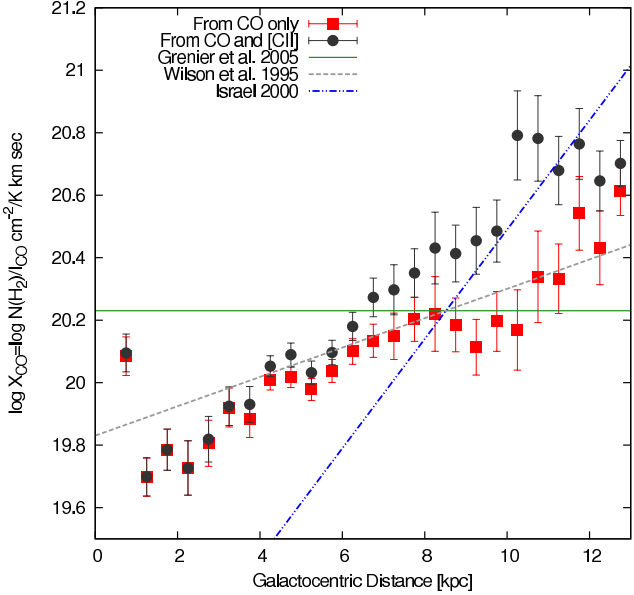}
      \caption{Radial distribution of the CO--to--H$_2$ conversion
      factor. We show the distribution for values estimated
      considering the CO--traced H$_2$ gas as well as that including the
      CO--dark H$_2$ gas  component.}
\label{fig:xcp}
\end{figure}

\subsection{The CO--to--H$_2$ conversion factor in the Milky Way}
\label{sec:co-h_2-conversion}

When $^{13}$CO observations are not available, the mass of molecular
clouds or the molecular content of entire galaxies is often estimated
from observations of the $^{12}$CO $J= 1 \to 0$ line applying an
empirically--derived CO--to--H$_2$ conversion factor ($X_{\rm CO}
\equiv N({\rm H}_2)/I_{\rm CO} \propto M_{{\rm H}_2}/L_{\rm
CO}$). Modelling applied to $\gamma$--ray observations, which trace
the total molecular content along the line--of--sight, results in a
local value of $1.74\times10^{20}$ cm$^{-2}$ (K km s$^{-1}$)$^{-1}$ or
$M_{\rm H_2}/L_{\rm CO}=3.7\,{\rm M}_{\odot}$(K Km s$^{-1}$
pc$^{2}$)$^{-1}$ \citep{Grenier2005}.  Several theoretical studies
have focused on the dependence of $X_{\rm CO}$ on environmental
conditions, showing that $X_{\rm CO}$ is particularly sensitive to
metallicity and the strength of the FUV radiation field
\citep[e.g.][]{MaloneyBlack88,Sakamoto96}. Observationally, however,
it has been found that $X_{\rm CO}$ is mostly sensitive to metallicity
\citep[e.g.][]{Rubio91,Wilson1995,Israel97,Israel00} while not showing
a significant dependence on the strength of the FUV field
\citep{Pineda2009,Hughes2010}. 

 We estimated the radial distribution
of $X_{\rm CO}$ in the Galactic plane by dividing the H$_2$ column
density in a given Galactocentric ring by its corresponding $^{12}$CO
emissivity.  In Figure~\ref{fig:xcp} we show the radial distribution
of $X_{\rm CO}$ derived from the H$_2$ column density traced by CO and
$^{13}$CO only, as well as that including the CO--dark H$_2$
component.  We also show the radial distribution of $X_{\rm CO}$
derived from the relationship between $X_{\rm CO}$ and metallicity
observed by \citet{Wilson1995} and \citet{Israel00} in nearby
galaxies. We converted the dependence of $X_{\rm CO}$ on metallicity
presented by these authors to a radial dependence using the slope of
the metallicity gradient in Equation~(\ref{eq:2}), fixing $X_{\rm CO}$
to be $1.74\times10^{20}$ cm$^{-2}$ (K km s$^{-1}$)$^{-1}$ at $R_{\rm
gal}=8.5$\,kpc.

Both $X_{\rm CO}$ distributions increase with the Galactocentric
distance. But when the contribution from CO--dark H$_2$ is included,
the slope is steeper.  This steeper slope is a result of the monotonic
increase of the CO--dark H$_2$ contribution to the total H$_2$ column
density with Galactocentric distance
(Figure~\ref{fig:nh2_fraction}). The radial distribution of $X_{\rm
CO}$ considering $N({\rm H}_2)$ from $^{12}$CO and $^{13}$CO only is
consistent with that fitted by \citet{Wilson1995}.  When CO--dark
H$_2$ is included in the calculation of $X_{\rm CO}$ the radial
distribution starts to resemble that fitted by \citet{Israel00}. The
offsets of course will change if the normalization at $R_{\rm
gal}$=8.5\,kpc is adjusted. Note that there is a difference in the
method used by \citet{Wilson1995} and \citet{Israel00} to derive
$X_{\rm CO}$. \citet{Wilson1995} observed $^{12}$CO in several giant
molecular clouds within several nearby galaxies and compared their
virial masses to the mass derived from their $^{12}$CO luminosities
using a Galactic conversion factor. \citet{Israel00} estimated the
H$_2$ mass of a sample of nearby galaxies by modelling their FIR
continuum emission to derive their masses and related them to their
$^{12}$CO luminosities to derive $X_{\rm CO}$. The FIR continuum
emission should trace the total H$_2$ content (including the CO--dark
H$_2$ component) while the method based on the virial mass derived
from $^{12}$CO mapping accounts for only the H$_2$ mass traced by
$^{12}$CO.
 
The increase of $X_{\rm CO}$ with Galactocentric distance for the case
when only CO and $^{13}$CO are used to derive $N({\rm H}_2)$ is a
result of the abundance gradient we applied in the calculation of
$N({\rm H}_2)$. If we set the [CO]/[H$_2$] abundance to be constant
when converting $N({\rm CO})$ to $N({\rm H}_2)$ for Mask CO2, and from
$A_{\rm V}$ to $N({\rm H}_2)$ for Mask\,CO1, we obtain a nearly
constant radial distribution of $X_{\rm CO}$. When the CO--dark H$_2$
is included, however, the slope (seen in Figure~\ref{fig:xcp}) results
from both the abundance and the thermal pressure gradients. If we also
set the [C$^+$]/[H$_2$] abundance to be constant we still obtain a
distribution that increases with $R_{\rm gal}$.  Thus, the difference
in the slopes in the distribution of $X_{\rm CO}$ is a result of the
use of the thermal pressure gradient in the calculation of the
CO--dark H$_2$ column density distribution.

\subsection{The surface density distribution and mass of the atomic and molecular gas in the Galactic plane }
\label{sec:surf-dens-distr}

\begin{figure}[t]
  \centering
   \includegraphics[width=0.45\textwidth]{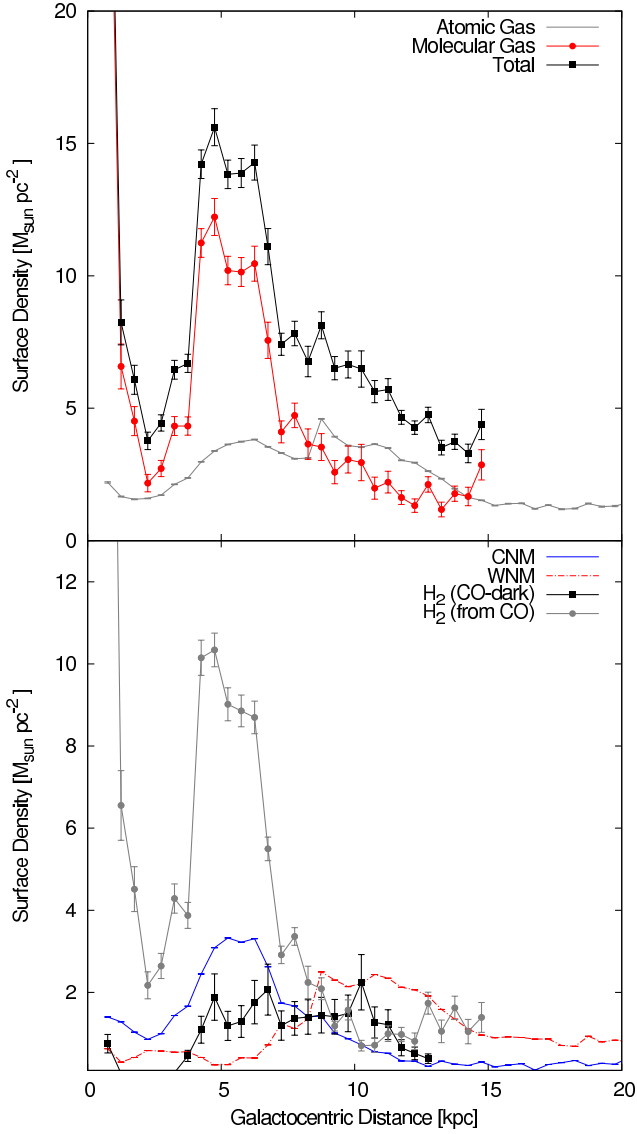}
      \caption{({\it top panel}) Radial distribution of the surface
 density of hydrogen in the Galactic plane. We also show the
 contribution of the atomic and molecular hydrogen gas. ({\it lower panel})
 Radial distribution of the different ISM components,
 including the cold and warm neutral medium (atomic gas), the CO--dark
 H$_2$ component, and the H$_2$ gas that is traced by $^{12}$CO and
 $^{13}$CO. All surface densities have been corrected to account for
 the contribution of He.}
\label{fig:distro_surf}
\end{figure}

In the top panel of Figure~\ref{fig:distro_surf}, we show the vertical
surface density distribution of the total hydrogen gas in the Galactic
plane. We also show the relative contributions from atomic and
molecular gas. We calculated the surface densities assuming a Gaussian
vertical distribution of the gas. For the molecular gas we assume a
constant FWHM disk thickness of 130\,pc
(Section~\ref{sec:radi-distr-plots}). In the case of the atomic gas,
the disk thickness is observed to increase with Galactocentric
distance, and we used the exponential relation presented by
\citet{Kalberla2009}, z$_{\rm FWHM}=0.3{\rm e}^{(R_{\rm
gal}-8.5)/9.8}$\,kpc.  In all cases, the surface density includes a
correction to account for the contribution from He. In the lower panel
of Figure~\ref{fig:distro_surf}, we show the surface density
distribution of the different contributing ISM components of the
atomic and molecular gas studied here, the cold and warm neutral
medium, the CO--dark H$_2$ gas, and the H$_2$ gas traced by $^{12}$CO
and $^{13}$CO.  The surface density distribution of atomic and
molecular gas derived in our analysis is similar to that derived by
\citet{Scoville1987}, but with the addition of the CO--dark H$_2$
component, the distribution of molecular gas extends over a larger
range of Galactocentric distances compared with that traced by
$^{12}$CO and $^{13}$CO only.  In Table~\ref{tab:mass}, we list the
mass of each ISM component studied here. The CO--dark H$_2$ gas
accounts for about 30\% of the total molecular gas mass. The WNM
dominates the mass of atomic gas in the Galactic plane, accounting for
70\% of the total atomic mass.

\begin{table} [t]                                           
\caption{Mass of  Milky Way ISM components}
\label{tab:mass}
\centering                                                  
\begin{tabular}{l c }                                 
\hline\hline
Component &  Mass \\
 &  $10^{9}$\,M$_{\odot}$      \\
\hline 
H$_2$ from CO &  1.9 \\
H$_2$ CO--dark & 0.7 \\
H$_2$ Total &    2.6 \\
\hline 
H\,{\sc i} CNM &   0.9 \\
H\,{\sc i} WNM &   1.6 \\
H\,{\sc i} Total & 2.5 \\
\hline
Total H\,{\sc i}+H$_2$ & 5.1 \\
\hline
\end{tabular}                                               
\end{table}

\begin{figure}[t]
  \centering
   \includegraphics[width=0.45\textwidth]{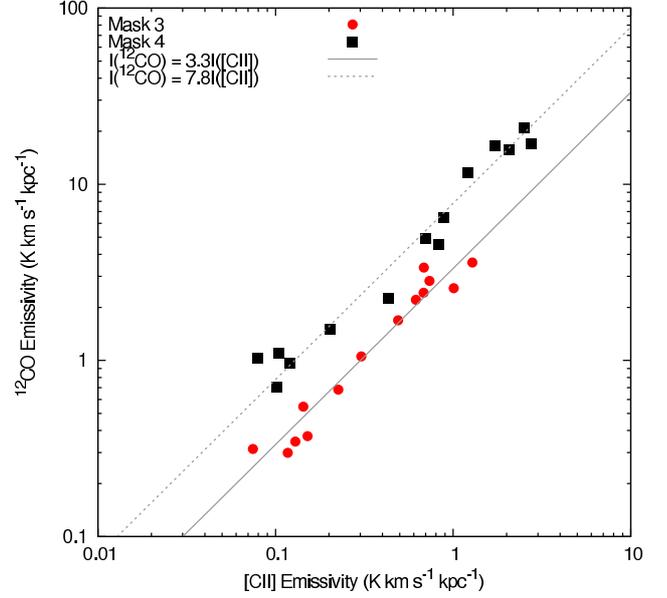}
      \caption{Comparison between the [C\,{\sc ii}] and $^{12}$CO
emissivities corresponding to different Galactocentric rings, for gas
associated with Mask\,3 (red circles) and Mask\,4 (black boxes). The
straight lines represent fits to the data. Typical uncertainties are
0.02 and 0.1 K\,km\,s$^{-1}$\,kpc$^{-1}$, for [C\,{\sc ii}] and
$^{12}$CO, respectively.}
\label{fig:intensity_comparison}
\end{figure}

\begin{figure}[t]
  \centering
   \includegraphics[width=0.45\textwidth]{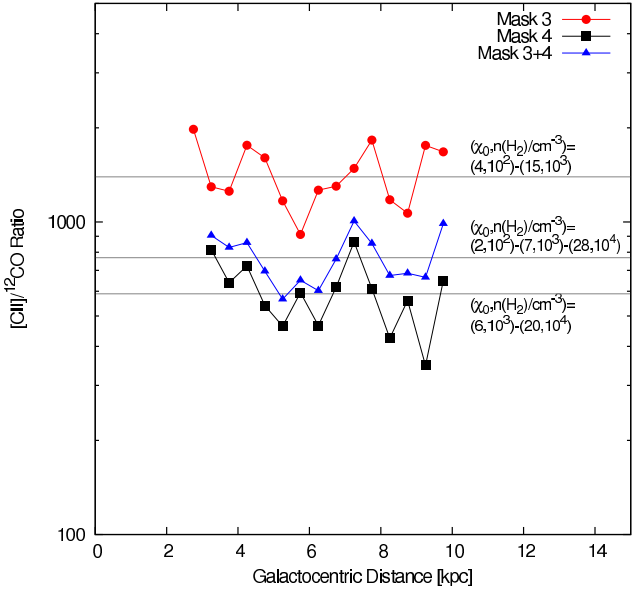}
      \caption{Radial distribution of the [C\,{\sc ii}] to $^{12}$CO
emissivity ratio for Mask\,3 (red circles) and Mask 4 (black boxes).
We also show the radial distribution of the [C\,{\sc ii}]/$^{12}$CO
ratio resulting from the combined emission from Mask 3 and 4 (blue
triangles). The horizontal lines correspond to the average value for
each Mask region between 3--10\,kpc.   The horizontal lines are
labeled with a range of pairs of the FUV field and H$_2$ volume
densities ($\chi_0$, $n_{\rm H_2}$), that reproduce their
corresponding [C\,{\sc ii}]/$^{12}$CO ratio, as predicted by the
\citet{Kaufman99} PDR model.   To facilitate the comparison with the
PDR model, the ratios were calculated using [C\,{\sc ii}] and
$^{12}$CO emissivities expressed in units of erg\,
s$^{-1}$\,cm$^{-2}$\,sr$^{-1}$\,kpc$^{-1}$. In these units, typical
uncertainties in the ratios are 180, 33, and 35 for Mask\,3, Mask\,4,
and the combined emission from both Mask regions, respectively.  }
\label{fig:ratio_distro}
\end{figure}

\section{The FUV radiation field distribution in the Milky Way}
\label{sec:fuv-radiation-field}

An important parameter governing the thermal balance and chemistry in
interstellar clouds is the strength of the FUV field, $\chi_0$ (in
units of the \citealt{Draine78} field). The gas heating in
cold, diffuse regions is dominated by the photoelectric effect
resulting from the absorption of FUV photons by dust grains and PAHs. The FUV
field also plays a critical role in the formation and destruction of
CO, determining the column density at which the C$^+$/C$^0$/CO
transition takes place. It is therefore of interest to determine the
typical strength of the FUV field to which the interstellar gas is
exposed and how this FUV field varies from place to place in the plane
of the Milky Way.

PDR model calculations suggest that the [C\,{\sc ii}]/$^{12}$CO
integrated intensity ratio is primarily a function of $\chi_0$ and the
total H ($n_{\rm HI}+2n_{\rm H_2}$) volume density \citep{Wolfire1989,
Kaufman99}. The reason for this dependence is that the $^{12}$CO
emission becomes optically thick quickly after a modest fraction of
the gas--phase carbon is converted to CO.  The [C\,{\sc ii}]/$^{12}$CO
ratio is thus determined by the column density of C$^+$ and the
temperature of the line--emitting region. Both quantities depend, in
turn, on the strength of FUV field and H volume density.  The solution
for a given value of the [C\,{\sc ii}]/$^{12}$CO ratio is not unique,
however, and several pairs of $\chi_0$ and $n_{\rm H}$ can reproduce
the observations. Therefore, we can only use the [C\,{\sc
ii}]/$^{12}$CO ratio to determine a range of plausible values of the
FUV radiation field and H volume density that characterize the
line--emitting region.

In a preliminary study of [C\,{\sc ii}], $^{12}$CO, and $^{13}$CO
velocity components on 16 GOT\,C+ LOSs, \citet{Pineda2010b} used PDR
model calculations to constrain the strength of the FUV field to be
mostly $\chi_0=$1--10, over a wide range of H$_2$ volume densities,
thus suggesting that most of the observed [C\,{\sc ii}] emission
associated with $^{12}$CO and $^{13}$CO is exposed to weak FUV
radiation fields. A similar result, was obtained from PDR modeling of
the [C\,{\sc ii}] emission observed by COBE by \citet{Cubick2008},
suggesting that most of the [C\,{\sc ii}] in our Galaxy originates in
a clumpy medium exposed to an FUV field of $\chi_0\simeq60$.

With the complete GOT\,C+ survey, we can study the [C\,{\sc
ii}]/$^{12}$CO ratio over the entire Galactic plane and determine how the
strength of the FUV field is distributed in the Galaxy. We consider
the azimuthally averaged emissivity of [C\,{\sc ii}] calculated from
spaxels where both $^{12}$CO and $^{13}$CO emission is detected (Mask
4), and where $^{12}$CO is detected but $^{13}$CO is not (Mask 3; See
Figure~\ref{fig:mask_radial_distro}). As discussed in
Section~\ref{sec:origin-c-sc}, gas associated with Mask\,3 might
represent low--column density gas surrounding the high--column density
regions associated with Mask\,4.  In
Figure~\ref{fig:intensity_comparison}, we show a comparison between
the [C\,{\sc ii}] and $^{12}$CO intensities for different
Galactocentric rings, for gas associated with Mask\,3 and 4. In both
mask regions we see a good correlation between these two
emissivities. A straight line fit results in $I({\rm
^{12}CO})=7.8I($[C\,{\sc ii}]), for Mask 3, and $I({\rm
^{12}CO})=3.3I($[C\,{\sc ii}]) for Mask 4.

In Figure~\ref{fig:ratio_distro}, we show the radial distribution of
the [C\,{\sc ii}] to $^{12}$CO emissivity ratio for Masks\,3 and 4.
We also include the radial distribution of the [C\,{\sc ii}]/$^{12}$CO
ratio resulting from the combined emission from Masks 3 and 4. The
ratio is computed considering emissivities in units of erg s$^{-1}$
cm$^{-2}$ sr$^{-1}$ kpc$^{-1}$.  We restrict the range in
Galactocentric distance to be between 3 and 10\,kpc because, for these
regions, both [C\,{\sc ii}] and $^{12}$CO have significant emission in
this range.  As expected, given the linear correlation between
[C\,{\sc ii}] and $^{12}$CO, the ratio shows little variation between
3 and 10\,kpc.   We also include horizontal lines representing the
average [C\,{\sc ii}]/$^{12}$CO ratio for each mask region. These
horizontal lines are labeled with a range of pairs of the FUV field
and H$_2$ volume densities ($\chi_0$, $n_{\rm H_2}$), that can
reproduce the corresponding values of the [C\,{\sc ii}]/$^{12}$CO
ratio, as predicted by the PDR model calculations from
\citet{Kaufman99}.  As mentioned before, Mask 3 might represent
regions with lower column densities compared to those in Mask\,4. This
column density difference might also translate into a difference in
volume densities because we do not expect a large difference in the
spatial extent of such regions. We thus show the results of the PDR
model for ranges in $n_{\rm H_2}$ that, based on the analysis in
Section~\ref{sec:h_2-gas-traced}, might be appropriate. We considered
the range of H$_2$ volume density $10^2-10^3$\,cm$^{-3}$ for Mask\,3,
and $10^3-10^4$\,cm$^{-3}$ for Mask\,4.  For the combined emission
from Mask\,3 and Mask\,4, we considered a wider range of possible
H$_2$ volume densities $10^2-10^4$\,cm$^{-3}$.  Note that the PDR
model calculations presented by \citet{Kaufman99} are in terms of the
total H volume density, including atomic and molecular hydrogen. For
the comparison between the observations and model predictions we
assumed that the contribution of atomic hydrogen to the total H volume
density is negligible and show the results of the PDR model in terms
of the H$_2$ volume density.

We find that over the selected range of H$_2$ volume densities, the
[C\,{\sc ii}]/$^{12}$CO ratio in both Mask\,3 and Mask\,4 suggest
values of $\chi_0$ that are in the range $1$--$30$, while we do not find
any systematic variation in $\chi_0$ with Galactocentric distance
between 3\,kpc and 10\,kpc. The same range of $\chi_0$ is
suggested by the [C\,{\sc ii}]/$^{12}$CO ratio resulting from the
combined emission from Mask\,3 and 4, over a wider range of H$_2$
volume densities.  Our results therefore suggest that on average the
[C\,{\sc ii}] emission associated with $^{12}$CO and/or $^{13}$CO
observed in the Galactic plane by the GOT C+ survey emerges from gas
exposed to modest FUV radiation fields, away from massive star--forming
regions.

\section{Summary}
\label{sec:conclusions}

  In this paper we have presented results of our study of the
  distribution and properties of the different ISM components in the
  plane of the Milky Way using spatially and velocity resolved
  observations of the [C\,{\sc ii}] 158$\mu$m line.  Our results can
  be summarized as follows:

\begin{enumerate}

\item We presented the first longitude--velocity maps of the [C\,{\sc
ii}] 158$\mu$m line in the plane of the Milky Way.   We found that the
[C\,{\sc ii}] emission is mostly associated with the spiral arms,
tracing the envelopes of evolved clouds as well as clouds that
are in the transition between the atomic and molecular phases.


\item We found that most of the [C\,{\sc ii}] emission in the Galactic
plane emerges from Galactocentric distances between 4\,kpc and
11\,kpc.

\item We estimated that most of the observed [C\,{\sc ii}] emission
emerges from dense photon dominated regions ($\sim$47\%), with  smaller
contributions from CO--dark H$_2$ gas ($\sim$28\%), cold atomic gas ($\sim$21\%),
and ionized gas ($\sim$4\%).

\item We used the [C\,{\sc ii}] emission to separate the WNM and CNM
contributions in the Galactic plane.  We find that most of the atomic gas inside the Solar
radius is in the form of cold neutral medium, while the warm neutral
medium gas dominates the outer Galaxy. The average fraction of the
total atomic gas that is CNM is $\sim$43\%.



\item We found that the warm and diffuse CO--dark H$_2$ is distributed
over a larger range of Galactocentric distances (4--11\,kpc) than the
cold and dense H$_2$ gas traced by $^{12}$CO and $^{13}$CO
(4--8\,kpc). The fraction of the total H$_2$ gas that is CO-dark H$_2$
increases with Galactocentric distance, ranging from $\sim$20\% at 4\,kpc up
to $\sim$80\% at 10\,kpc.  We estimate that the CO--dark H$_2$ gas
component accounts for $\sim$30\% of the total molecular gas mass in the
Milky Way.

\item We estimated the radial distribution of the CO--to--H$_2$
conversion factor.  When $N({\rm H}_2)$ is derived from 
$^{12}$CO and $^{13}$CO only, the CO--to--H$_2$ conversion factor
increases with Galactocentric distance, influenced by the Galactic
metallicity gradient. When the contribution of CO--dark H$_2$ gas is
included in $N({\rm H}_2)$, the CO--to--H$_2$ conversion factor
increases with Galactocentric distance with an steeper slope, which is
influenced by both  metallicity and  thermal pressure gradients.

\item We estimated the average strength of the FUV field modelling the
[C\,{\sc ii}]/$^{12}$CO ratio. We find that most of the observed
[C\,{\sc ii}] emission emerging from dense PDRs is associated with modest FUV
fields in the $\chi_0\simeq1-30$ range.

\end{enumerate}   

Our results confirm that [C\,{\sc ii}] is an important tool to
characterize the different components of the ISM. Large--scale mapping
of the Milky Way and nearby galaxies with orbital and sub--orbital
observatories are essential to further characterize the life-cycle of
the interstellar medium, and will help us to build important templates
for the interpretation of observations of [C\,{\sc ii}] at
high--redshift.

\begin{acknowledgements}
 
This research was conducted at the Jet Propulsion Laboratory,
California Institute of Technology under contract with the National
Aeronautics and Space Administration.  We thank the staffs of the ESA
and NASA Herschel Science Centers for their help.  We would like to
thank Roberto Assef for enlightening discussions.  The Galactic
Arecibo L-Band Feed Array H\,{\sc i} (GALFA--H\,{\sc i}) Survey data
set was obtained with the Arecibo L-band Feed Array (ALFA) on the
Arecibo 305m telescope.  Arecibo Observatory is part of the National
Astronomy and Ionosphere Center, formerly operated by Cornell
University under Cooperative Agreement with the National Science
Foundation of the United States of America. \copyright\ 2013 California
Institute of Technology. Government sponsorship acknowledged.

\end{acknowledgements}

\bibliographystyle{aa} 
\bibliography{papers}

\appendix

\section{Azimuthally averaged line emissivities}
\label{sec:determ-azim-aver}

  \begin{figure*}[t]
   \centering
   \includegraphics[width=0.75\textwidth]{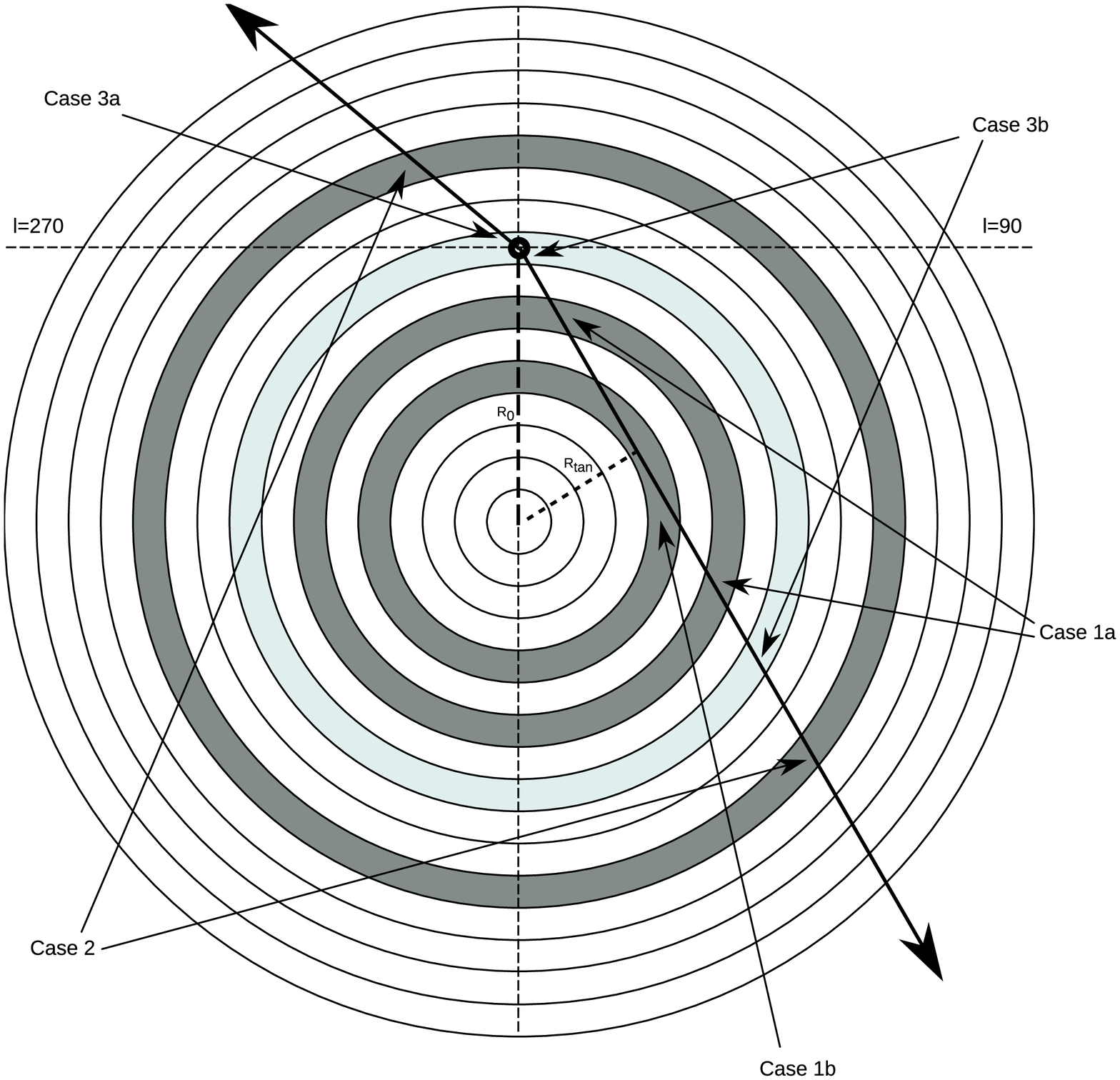}
      \caption{Schematic of the Milky Way divided by a set of rings
       centered at the Galactic center. The two thick arrows
       originating from the Sun (shown as a circle) correspond to two
       lines--of--sight, one looking toward the inner Galaxy and the
       other toward the outer Galaxy. The dashed lines represent the
       distance to the Galactic center ($R_{\odot}$) and the distance to the
       tangent point ($R_{\rm tan}$) for the line of sight that goes
       towards the inner Galaxy. The thin lines point to the different
       path lengths crossed by the lines--of--sight that correspond to
       the different cases considered in the text.}
\label{fig:Schematic}
   \end{figure*}

  \begin{figure}[t]
   \centering
   \includegraphics[width=0.48\textwidth]{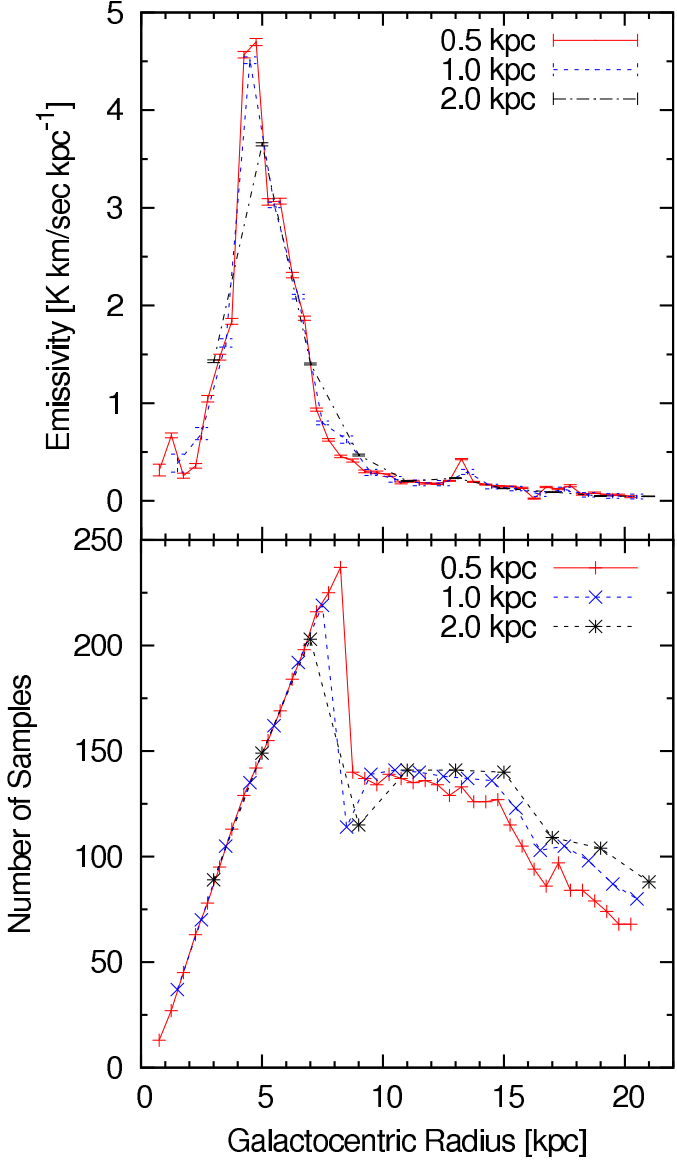}
      \caption{({\it top}) Radial distribution of the [C\,{\sc ii}]
emissivity for different values of the radial extent for each ring
$\Delta R= 0.5$\,kpc, 1\,kpc, and 2\,kpc. ({\it bottom}) The number of
samples crossed by the GOT C+ lines--of--sight as a function of 
Galactocentric radius, for different values of $\Delta R$. }
\label{fig:different_bins}
   \end{figure}

In the following we describe the procedure we used to calculate the
radial distribution of the azimuthally averaged integrated intensity
per unit length of a spectral line. Assuming a flat rotation curve and
circular motions, the distance to the Galactic center $R$ for a given
data point with Galactic longitude $l$, latitude $b$, and local
standard of rest (LSR) velocity $V_{\rm LSR}$, is given by

\begin{equation}
\label{eq:9}
R= R_{\odot} \left ( 1+\frac {V_{\rm LSR}}{V_{\odot}\sin(l)\cos(b)}
\right )^{-1},
\end{equation}
where $R_{\odot}$ is the distance from the sun to the Galactic center
and $V_{\odot}$ is the orbital velocity of the sun with respect to the
Galactic center. We adopt the IAU recommended values of $R_{\odot}$ =
8.5\,kpc and $V_{\odot}$ = 220 km s$^{-1}$. We can invert
Equation~(\ref{eq:9}) to relate the observed LSR velocity of the
emission to the Galactocentric distance through

\begin{equation}
\label{eq:10}
V_{\rm LSR} = V_{\odot}\sin(l)\cos(b)\left (\frac{R_{\odot}}{R}-1 \right ).
\end{equation}

We divided the Galaxy in a set of concentric rings, with a central
radial distance $R$ and a radial extent $\Delta R=R_{\rm out}-R_{\rm
in}$.  For a given ring, we used Equation~(\ref{eq:10}) to calculate
the corresponding interval limits in $V_{\rm LSR}$. We then calculated
the integrated intensity $I(l,R)$ within this LSR velocity range. The
integrated intensity depends on the path length crossed by a given
line of sight, $L(l,R_{\rm out},R_{\rm in})$, and therefore we divided
by this path length to calculate the integrated intensity per unit
length (emissivity).  The units of the emissivity are K km s$^{-1}$
kpc$^{-1}$.  The value of $L$ depends on the location of the Galaxy
we are sampling. This dependency is illustrated in
Figure~\ref{fig:Schematic} where we show a schematic of the different
cases we considered for the calculation of $L$.

In Case 1a we considered rings with outer radii inside the solar
circle ($R_{\rm out}\leq R_{\odot}$) and inner radius $R_{\rm in}$ larger
than the radius of the tangent $R_{\rm tan}=|R_{\odot}\sin(l)|$. In this
case a given ring is crossed twice by a given LOS. The path length for
one of these crossings is given by,

\begin{equation}
\label{eq:11}
L=\sqrt{(R_{\rm out})^2-(R_{\rm tan})^2}-\sqrt{(R_{\rm in})^2-(R_{\rm tan})^2}.
\end{equation}

In Case 1b we considered rings with $R_{\rm in}<R_{\rm tan}\leq R_{\rm
 out}$. In this case a ring is crossed only once and the path length
 is given by,

\begin{equation}
\label{eq:12}
L=2\sqrt{(R_{\rm out})^2-(R_{\rm tan})^2}.
\end{equation}

In Case 2 we considered LOSs that cross rings in the outer Galaxy
with $R_{\rm in}>R_{\odot}$. These rings are crossed once and the path length is
given by Equation~(\ref{eq:11}).

Rings in Case 3a have $R_{\rm in}<R_{\odot}<R_{\rm out}$ and are
observed in LOSs in the range $270$\degr $> l >90$\degr. In this case
the path length is  given by the distance to the point with $R=R_{\rm
out}$,

\begin{equation}
\label{eq:13}
L=R_{\odot}\cos(l)+\sqrt{(R_{\rm out})^2-(R_{\rm tan})^2}.
\end{equation}

Finally for Case 3b, we considered rings with $R_{\rm
in}<R_{\odot}<R_{\rm out}$ observed in LOSs in the range $270$\degr $<
l <90$\degr. In this case we cross a ring twice. For the near side the
path length is given by

\begin{equation}
\label{eq:14}
L=R_{\odot}\cos(l)+\sqrt{(R_{\rm in})^2-(R_{\rm tan})^2},
\end{equation}
and for the far side is given by Equation~(\ref{eq:11}).

The final step is to average the emissivities over all samples of a
given ring. This sum is done over all rings sampled by the GOT C+
survey $N_s$,

\begin{equation}\label{eq:5}
I_0(R)=\frac{\sum^{N_s}_i I_i/L_i }{N_s}.
\end{equation}

We estimated the errors on the emissivity by propagating the errors of
the individual determinations of the integrated intensity per unit
length, $\Delta I/L$, using,

\begin{equation}\label{eq:5}
\Delta I_0(R)=\frac{\sqrt{\sum^{N_s}_i (\Delta I_i/L_i)^2}}{N_s}.
\end{equation}

 In the top panel of Figure\,\ref{fig:different_bins} we show the
resulting radial distribution of the [C\,{\sc ii}] emissivity for
different values of $\Delta R= 0.5$\,kpc, 1\,kpc, and 2\,kpc. We can
see that for larger values of $\Delta R$ the peak shifts to around
5\,kpc, which is the midpoint where most of the [C\,{\sc ii}] emission
observed in the GOT C+ survey arises in the Galaxy. We adopt $\Delta
R= 0.5$\,kpc to preserve the details of the [C\,{\sc ii}] emissivity
distribution. The uncertainty in the [C\,{\sc ii}] emissivities for
this selection of $\Delta R$ is typically 0.02 K
km\,s$^{-1}$\,kpc$^{-1}$.  The bottom panel shows the number of samples
($N_s$) for a given ring. As we can see, rings with radii below 2\,kpc
are undersampled in our survey. Given that such rings are sampled by
lines--of--sights with $|l|<3$\degr, and that these LOSs are also
known to have non-circular motions, we exclude these LOSs in our
analysis. This exclusion has a minimal effect on emissivity
distribution beyond 2\,kpc.

\section{Position--velocity maps}
\label{sec:posit-veloc-maps-4}

\begin{figure*}[h]
   \centering
   \includegraphics[width=0.865\textwidth]{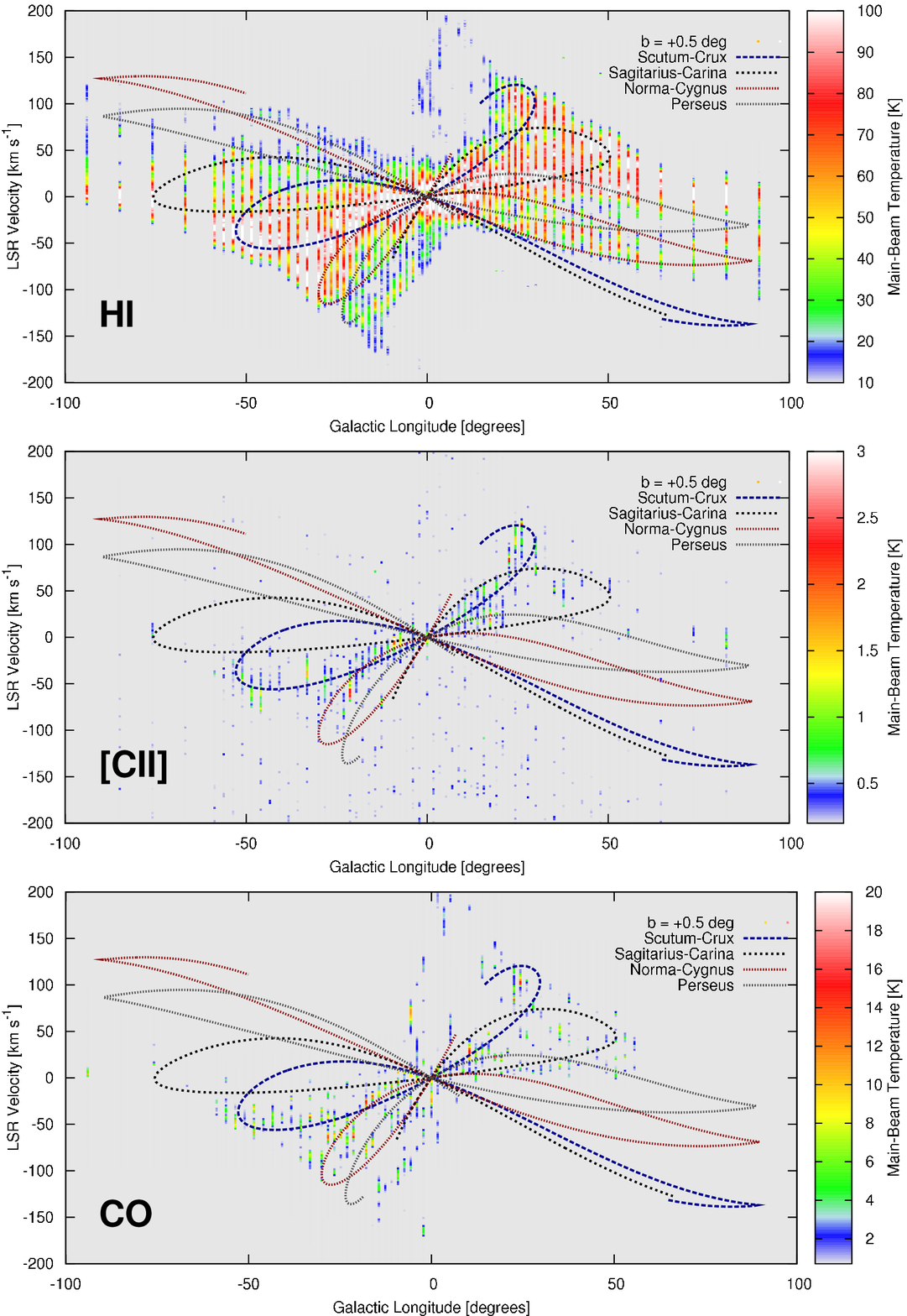}
      \caption{Position--velocity maps of the Milky Way in [C\,{\sc
              ii}] observed by GOT\,C+ for $b=+0.5$\degr. }
   \label{fig:pvmap1}
   \end{figure*}

\begin{figure*}[h]
   \centering
   \includegraphics[width=0.865\textwidth]{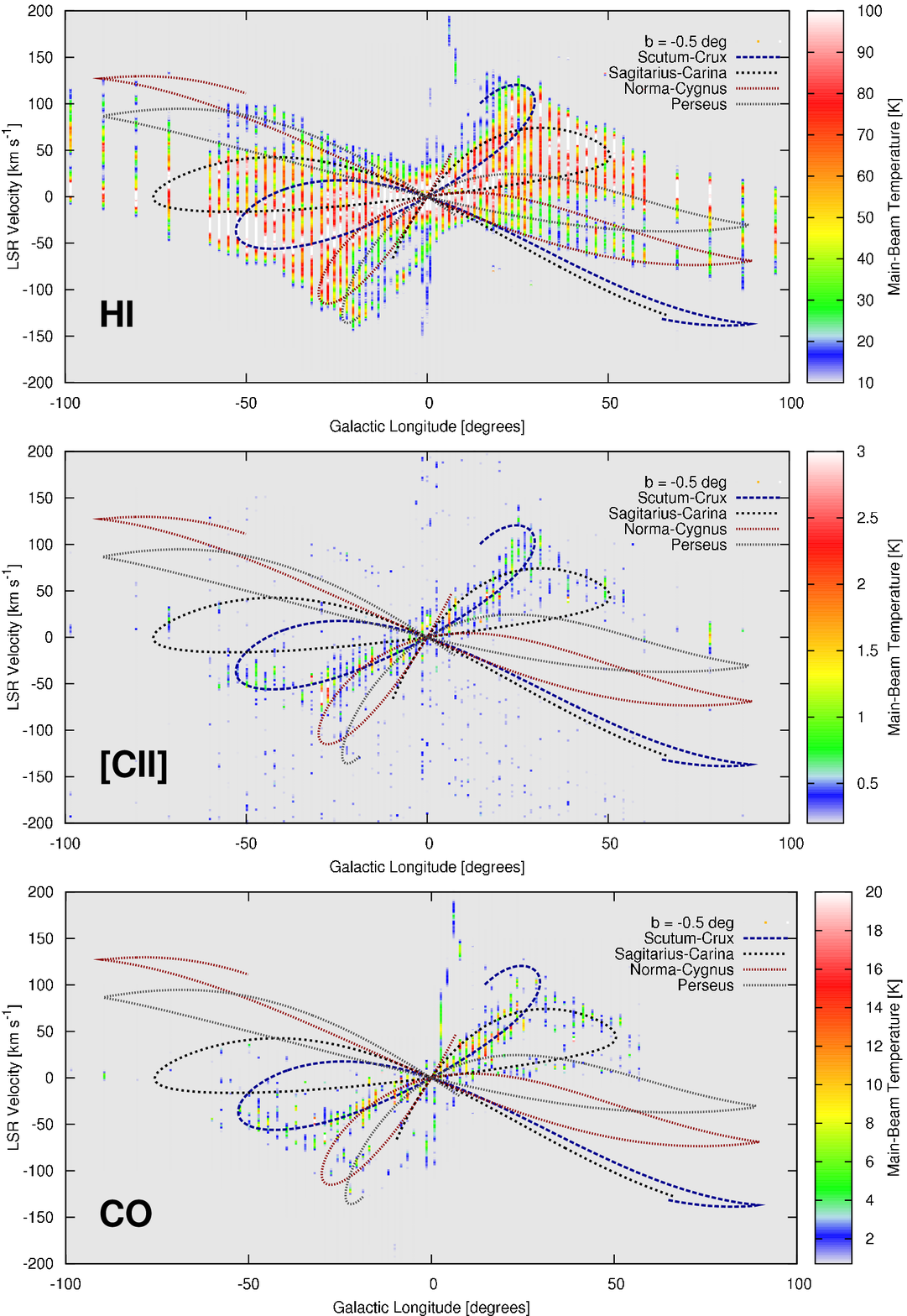}
      \caption{Position--velocity maps of the Milky Way in [C\,{\sc
              ii}] observed by GOT\,C+ for $b=-0.5$\degr. }
   \label{fig:pvmap2}
   \end{figure*}

\begin{figure*}[h]
   \centering
   \includegraphics[width=0.865\textwidth]{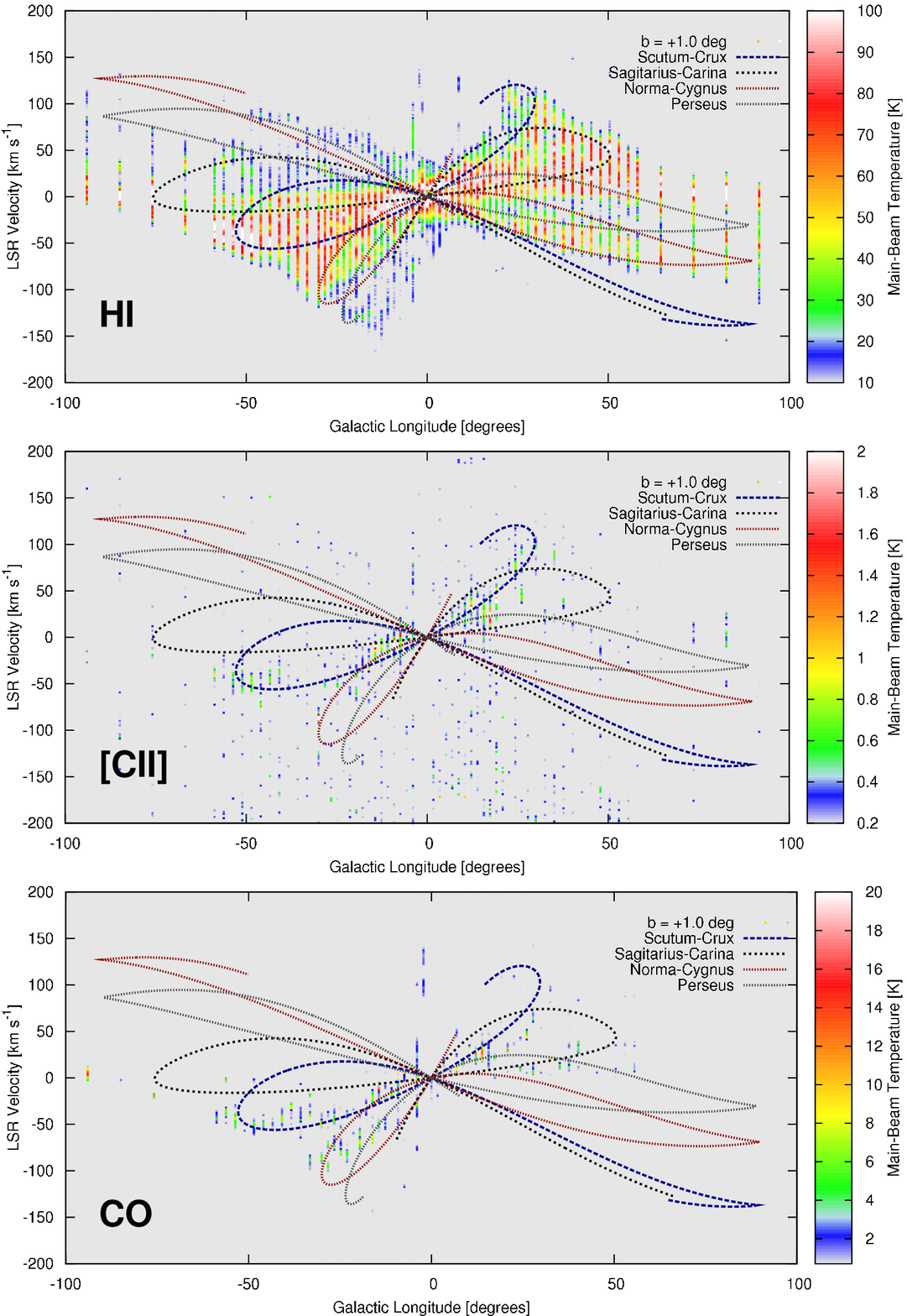}
      \caption{Position--velocity maps of the Milky Way in [C\,{\sc
              ii}] observed by GOT\,C+ for $b=+1.0$\degr. }
   \label{fig:pvmap3}
   \end{figure*}

\begin{figure*}[h]
   \centering
   \includegraphics[width=0.865\textwidth]{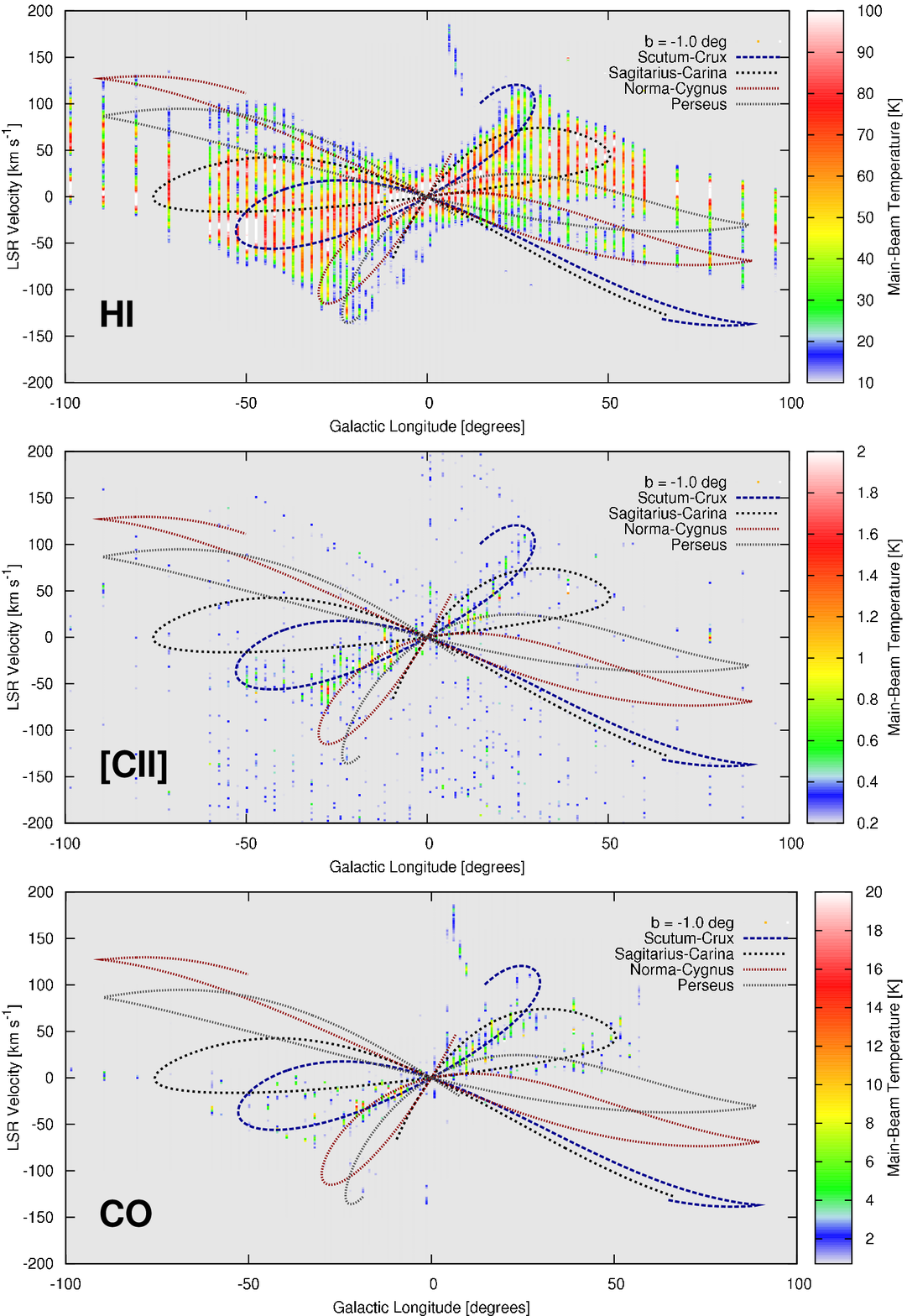}
      \caption{Position--velocity maps of the Milky Way in [C\,{\sc
              ii}] observed by GOT\,C+ for $b=-1.0$\degr. }
   \label{fig:pvmap4}
   \end{figure*}

\begin{figure*}[h]
   \centering
   \includegraphics[width=0.65\textwidth]{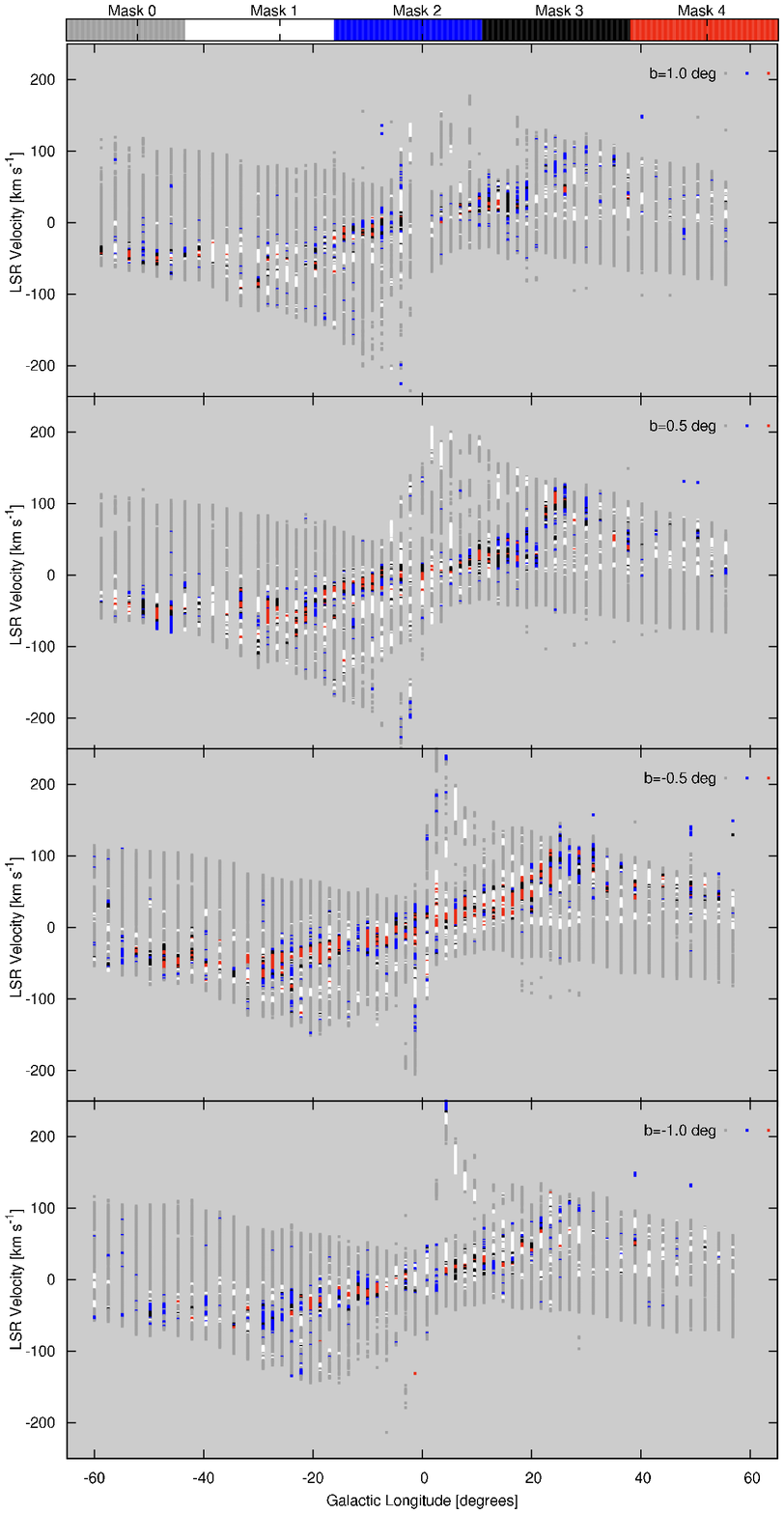}
      \caption{Position velocity maps of the different cloud types
              defined in Section~\ref{sec:velocity-components} for
              $b=\pm0.5$\degr\ and $\pm1.0$\degr. The Mask 0 ({\it
              grey}) represents velocity components with only H\,{\sc
              i} detected, Mask 1 ({\it white}) are components with
              only H\,{\sc i} and CO detected, Mask 2 ({\it blue})
              components with only H\,{\sc i} and [C\,{\sc ii}], Mask
              3 ({\it black}) components with H\,{\sc i}, [C\,{\sc
              ii}], and CO, and Mask 4 ({\it Red}) components with
              H\,{\sc i}, [C\,{\sc ii}], $^{12}$CO, and $^{13}$CO.}
\label{fig:mask2}
   \end{figure*}

In Figures~\ref{fig:pvmap1} to ~\ref{fig:pvmap4}, we present the
position--velocity maps of the [C\,{\sc ii}], CO, and H\,{\sc i} emission
for the GOT\,C+ lines--of--sight with $b=\pm0.5$\degr\ and
$b=\pm1.0$\degr. Figure~\ref{fig:mask2} shows the position--velocity maps
for the different cloud types discussed in
Section~\ref{sec:velocity-components}, for the GOT\,C+ lines--of--sight
with  $b=\pm0.5$\degr\ and
$b=\pm1.0$\degr.


\end{document}